 \documentclass[twocolumn,aps,pra,superscriptaddress,showpacs]{revtex4-1}
\usepackage{bm}
\usepackage{color}
\usepackage{amsmath}
\usepackage{amssymb}
\usepackage{graphicx}

\definecolor{gray}{cmyk}{0,0,0,.5}

\definecolor{darkblue}{cmyk}{1,1,0,0.1}
\definecolor{lightblue}{rgb}{.66,.66,1}
\definecolor{lightcyan}{cmyk}{.5,0,0,0}
\definecolor{lightred}{cmyk}{0,.7,.8,0}
\definecolor{darkred}{cmyk}{0,1,1,0.1}

\newcommand{\bra}[1]{\langle{#1}|}
\newcommand{\ket}[1]{|{#1}\rangle}

\newcommand{\braopket}[3]{\langle #1| #2 | #3\rangle}

\newcommand{\B}[1]{$B=#1\,$G}
\newcommand{\G}[1]{$G=#1\,$Tm$^{-1}$}
\newcommand{\F}[1]{$F=10^{-#1}\,${a.u.}}

\usepackage{nicefrac}
\newcommand{\cm}{{center of mass}} 
\newcommand{\au}{{a.u.}} 
\newcommand{\half}{ \text{\nicefrac{1}{2}} }

\begin{document}

\title{Interaction-induced stabilization of circular Rydberg atoms}
\author{Bernd Hezel}
\altaffiliation[Present address: ]{Potsdam Institute for Climate Impact Research, 14412 Potsdam, Germany} 
\affiliation{Physikalisches Institut, Universit\"at Heidelberg, Philosophenweg 12, 69120 Heidelberg, Germany}
\author{Michael Mayle}
\altaffiliation[Present address: ]{JILA, University of Colorado and National Institut of Standards and Technology, Boulder, Colorado 80309-0440, USA}
\affiliation{Zentrum f\"ur Optische Quantentechnologien, Universit\"at Hamburg, Luruper Chaussee 149, 22761 Hamburg, Germany}
\author{Peter Schmelcher}
\email[]{peter.schmelcher@physnet.uni-hamburg.de}
\affiliation{Zentrum f\"ur Optische Quantentechnologien, Universit\"at Hamburg, Luruper Chaussee 149, 22761 Hamburg, Germany}

\date{\today}

\begin{abstract}
We discuss a candidate solution for the controlled trapping and manipulation of two individual Rydberg atoms by means of a magnetic Ioffe-Pritchard trap that is superimposed by a constant electric field. In such a trap 
Rydberg atoms experience a permanent electric dipole moment that can be of the order of several hundred Debye. 
The interplay of electric dipolar repulsion and three dimensional magnetic confinement 
leads to a well controllable equilibrium configuration 
with tunable trap frequency and atomic distance.
We thoroughly investigate the trapping potentials and analyze the interaction-induced stabilization of two such trapped Rydberg atoms. Possible limitations and collapse scenarios are discussed.
\end{abstract}

\pacs{
32.80.Ee, % Rydberg states
32.10.Ee, % Magnetic bound states, magnetic trapping of Rydberg states
32.60.+i, % Zeeman and Stark effects
37.10.Gh  % Atom traps and guides
}% PACS, the Physics and Astronomy Classification Scheme.
%\keywords{}

\maketitle
%\tableofcontents

\section{Introduction}
Among the many fascinating systems encountered in ultracold atomic and molecular physics 
are Rydberg atoms, i.e., highly excited atoms with large principal quantum number $n$. Their size can easily exceed that of ground state atoms by several orders of magnitude and at the same time is the origin of many extraordinary properties such as their massively enhanced response to external fields and, therewith, for their enormous polarizability \cite{gallagher94}. In ultracold gases, the resulting strong dipole-dipole interaction among Rydberg atoms has been found to give rise to a non-linear excitation behavior: Rydberg atoms strongly inhibit excitation of their neighbors entailing a state dependent local excitation blockade \cite{singer,tong,Liebisch2005,Vogt2007,Ditzhuijzen2008}, which on its part results in a collective excitation of many atoms \cite{Heidemann2007,Reetz-Lamour2008,Johnson2008}. Two recent experiments even demonstrated the blockade between two single atoms a few micrometers apart \cite{Urban2009,Gaetan2009}. From an application oriented point of view, the strong dipole blockade effect renders Rydberg atoms promising candidates for quantum information processing \cite{RevModPhys.82.2313} and allows the determination of the  interaction potential of Rydberg atoms in a one-dimensional lattice \cite{Mayle2011}. The large size of Rydberg atoms can also give rise to bonding interactions between Rydberg and ground state atoms. The scattering-induced attractive interaction binds the ground state atom to the Rydberg atom at a well-localized position within the Rydberg electron wave function and thereby yields giant ultra long-range molecules that can have internuclear separations of several thousand Bohr radii \cite{greene}. The spectroscopic characterization of such exotic molecular states, named trilobite and butterfly states on account of their particular electronic density, has succeeded recently \cite{Bendkowsky2009} and has triggered a revived theoretical and experimental activities \cite{PhysRevLett.105.163201,Butscher2010,PhysRevLett.104.243002,Rittenhouse2011a}.

Most of the experiments with Rydberg atoms still involve a large ensemble of atoms. They can therefore unavoidably solely investigate effective and averaged properties since individual atoms are typically not resolved. It is hence of great interest to study only a small number of Rydberg atoms that are preferably individually controllable and arrangeable with respect to one another. It is furthermore necessary to stabilize these Rydberg atom configurations against autoionization. An essential step in this direction is the trapping of electronically highly excited atoms.
Several works have focused on trapping Rydberg atoms, based on electric \cite{Hyafil2004}, optical \cite{Dutta2000,1367-2630-12-2-023031,PhysRevLett.104.173001}, or magnetic fields \cite{choi,LesanovskyPRL2005,Hezel2006,mayle:053410}. Due to the high level density and the strong spectral fluctuations with spatially varying fields, trapping or manipulation in general is a delicate task. This is particularly the case when both the center of mass and the internal motion are of quantum nature and the inhomogeneous external fields lead to an inherent coupling of these motions. 

In the present work we provide a candidate solution for the controlled trapping and manipulation of two individual Rydberg atoms by means of a magneto-electric trap. Specifically, we consider a magnetic Ioffe-Pritchard trap that is superimposed by a constant electric field which induces a permanent electric dipole moment for the Rydberg atoms that can be of the order of several hundred Debye. As has been shown in a previous work, the resulting dipole-dipole interaction in conjunction with the tight radial confinement of the Ioffe-Pritchard trap gives rise to an effectively one-dimensional ultracold Rydberg gas with a macroscopic interparticle distance \cite{mayle:113004}. Here, we consider in addition the longitudinal confinement that arises for a non-Helmholtz configuration of the Ioffe-Pritchard trap. In contrast to our previous work focusing on the trapping of individual Rydberg atoms in two dimensions \cite{LesanovskyPRL2005,Hezel2006,mayle:053410}, this allows the controlled confinement of two single Rydberg atoms in three dimensions with variable trapping parameters and distance. We thoroughly investigate the resulting trapping potentials and analyze the interaction-induced stabilization of two such trapped Rydberg atoms. Possible limitations and collapse scenarios are discussed.

In detail, we proceed as follows. In Section \ref{section:singleatom} the Hamiltonian of a single Rydberg atom in the magnetic Ioffe-Pritchard trap superimposed by a constant electric field is derived and the corresponding adiabatic potential surfaces for the center of mass motion of the Rydberg atom are provided, as well as analytic expressions for the electric dipole moment of the Rydberg atom induced by the external electric field. In Section \ref{s:iis} we consider the interaction of two Rydberg atoms in the same trapping environment. In the regime of strong transversal confinements, analytic expression for the equilibrium configuration of the two atoms are derived. Weakening this restriction leads to three-dimensional potential surfaces and possible loss mechanisms that are investigated in the remainder of the section. Section \ref{section:exschemes} outlines several routes to experimentally realize the proposed system. A brief summary is provided in Sec.~\ref{section:summary}. In the Appendix we present a detailed derivation of the perturbative results for the induced dipole moment of the Rydberg atoms in the considered trap.

%%%%%%%%%%%%%%%%%%%%%%%%%%%%%%%%%%%%%%%%%%%%%%%%%%%%%%%%%%%%%%%%%%%%%%%%%%%%%%%%%%%%%%%%%%%%%% 
%%%%%%%%%%%%%%%%%%%%%%%%%%%%%%%%%%%%%%%%%%%%%%%%%%%%%%%%%%%%%%%%%%%%%%%%%%%%%%%%%%%%%%%%%%%%%%
\section{Three-dimensional Ioffe-Pritchard confinement for a Rydberg atom\label{section:singleatom}}
%%%%%%%%%%%%%%%
\subsection{Two-body Hamiltonian for a single alkali Rydberg atom\label{ss:2bH}}
%\begin{itemize}
In a highly anisotropic magnetic field configuration like that of a Ioffe-Pritchard trap, the strength of the magnetic field can vary significantly over the extension of a Rydberg atom. The large size of Rydberg atoms can therefore modify the way they couple to the magnetic field compared to the coupling of ground state atoms. We incorporate the large extension of the atoms into our description by modeling a Rydberg atom by two particles, namely, a valence electron (particle 1) and an ionic core (particle 2). This is particularly appropriate for alkali atoms that are commonly used in Rydberg experiments. We include into our model the coupling of the electronic and the center of mass motion of the atom and hence do not resort to the infinitely heavy mass approximation. While the inclusion of the fine-structure and quantum defects can be readily done, it turns out not to be necessary for high angular momentum electronic state which we will be focusing on \cite{HezelPRA}. The coupling of the charged particles to the external magnetic field is introduced via the substitution, $\bm{p}_i\rightarrow \bm{p}_i-q_i\bm{A}(\bm{r}_i)$; $q_i$ is the charge of the $i$-th particle and $\bm{A}(\bm{x})$ is the vector potential belonging to the magnetic field $\bm{B}(\bm{x})$. Including the coupling of the magnetic moments due to the spins to the external field, our initial Hamiltonian in the laboratory frame reads (atomic units are used except when stated otherwise)
\begin{align} \label{e:Hinit}
   H_L={}&\frac{1}{2M_1}[\bm{p}_1-q_1\bm{A}(\bm{r}_1)]^2
   +\frac{1}{2M_2}[\bm{p}_2-q_2\bm{A}(\bm{r}_2)]^2 \nonumber\\
   &{} +V(\left|\bm{r}_1-\bm{r}_2\right|)
   -\bm{\mu}_1\cdot\bm{B}(\bm{r}_1)-\bm{\mu}_2\cdot\bm{B}(\bm{r}_2)
   \; .
 \end{align}
The magnetic moments of the particles are connected to the electronic spin $\mathbf{S}$ and the nuclear spin $\mathbf{\Sigma}$ according to $\mbox{\boldmath$\mu$}_1=-\mathbf{S}$ and $\mbox{\boldmath$\mu$}_2=-\frac{g_N}{2M_2}\mathbf{\Sigma}$, with $g_N$ being the nuclear $g$-factor; because of the large nuclear mass, the term involving $\mbox{\boldmath$\mu$}_2$ is neglected in the following.

The vector potential and the magnetic field of the Ioffe-Pritchard configuration read
\begin{align}           \label{eq:overallpotentialandfield}
    \bm{A} &= \underbrace{ \frac{B}{2} \left(\!\! \begin{array}{ccc} -y \\ x \\ 0 \end{array} \!\!\right)}_{=\bm{A}_c}
    +\underbrace{G\left(\!\! \begin{array}{ccc} 0 \\ 0 \\ x y \end{array} \!\!\right)}_{=\bm{A}_l}
    +\underbrace{\frac{Q}{4}\left(\!\!\begin{array}{ccc} 
          y(x^2+y^2-4 z^2) \\           -x(x^2+y^2-4 z^2) \\ 0 
        \end{array} \!\!\right)}_{=\bm{A}_q}
    , \nonumber\\ 
   \bm{B} &= \underbrace{B \left(\! \begin{array}{ccc} 0 \\ 0 \\ 1 \end{array} \!\right)}_{=\bm{B}_c}
    +\underbrace{G \left(\! \begin{array}{ccc} x \\ -y \\ 0 \end{array} \!\right)}_{=\bm{B}_l}
    +\underbrace{Q \left(\! \begin{array}{ccc} 
          -2xz \\           -2yz \\            -x^2-y^2+2 z^2
        \end{array} \!\right)}_{=\bm{B}_q}.
 \end{align}
The ``traditional'' macroscopic realization of the Ioffe-Pritchard trap uses four parallel current carrying Ioffe bars which generate the two-dimensional quadrupole field $\bm{B}_l$ that depends on the field gradient $G$. Encompassing Helmholtz coils create the additional constant field $\bm{B}_c$ where $B$ denotes the Ioffe field strength \cite{PhysRevLett.51.1336}. $\bm{B}_q$ designates the quadratic term generated by the Helmholtz coils whose magnitude, compared to the first Helmholtz term, can be varied by changing the geometry of the trap, 
\begin{equation}
  Q=B\cdot\frac{3}{2}\frac{4D^2-R^2}{(D^2+R^2)^{2}} =: B\cdot \tilde{Q}(D,R).
\end{equation}
$R$ is the radius of the Helmholtz coils, and $2D$ is their distance from each other. The geometry factor $\tilde{Q}(D,R)$ vanishes for $2D=R$, which is known as the Helmholtz configuration. Exposing Rydberg atoms to a Ioffe-Pritchard trap in a Helmholtz configuration has been extensively studied in Ref.~\cite{HezelPRA}. Here, we assume $\tilde{Q}$ to be non-zero and positive, $2D>R$. In this case, the absolute value of the magnetic field on the $Z$-axis, $|\bm B(0,0,Z)|=B|1+2\tilde{Q}Z^2|$, increases quadratically with $|Z|$. The geometry factor reaches its maximal value, $\tilde{Q}_{\text{max}}=\frac{9}{10}D^{-2}$, when $2D=\sqrt{6}R$, i.e., the smaller $R$, the larger $\tilde{Q}$.

Along the lines of Ref.~\cite{HezelPRA} we employ the unitary transformation $U=\exp\left\{\frac{i}{2}\,\bm{B}_c\times\bm{r}\cdot\bm{R}\right\}$, introduce relative and center of mass coordinates ($\bm{r}=\bm{r}_1-\bm{r}_2$ and $\bm{R}=(M_1 \bm{r}_1+M_2\bm{r}_2)/M$ with the total mass $M=M_1+M_2$), and omit the diamagnetic contributions. The Hamiltonian describing the Rydberg atom in the Ioffe-Pritchard trap becomes
\begin{eqnarray} \label{e:UHUQini}
     H  &=& \frac{\bm{P}^2}{2M}
    +H_A 
    + \frac{1}{2}\bm{L}_{\bm r}\cdot\bm{B}_c 
    + \bm{S}\cdot\bm B(\bm R + \bm r) 
    \nonumber\\
    &&  + \bm{A}_l(\bm{R}+\bm{r})\cdot\bm{p} 
    + \bm{A}_q(\bm{R}+\bm{r})\cdot\bm{p},
 \end{eqnarray}
where we named the electronic hydrogenic Hamiltonian in field-free space $H_A={\bm{p}^2}/{2} -{1}/{\bm r}$. In Hamiltonian (\ref{e:UHUQini}) we neglect any contributions arising from the transformation of $\bm{A}_q(\bm{R}+\bm{r})$ which is justified as long as $|X|,|Y|\ll \frac{2}{Bn}$. As usual, the principal quantum number of the Rydberg state is denoted as $n$. 

For all relevant laboratory field strengths the spectrum of the Hamiltonian (\ref{e:UHUQini}) is dominated by the field-free energies $E_A^{n}=-1/2n^2$ that are $n^2$-fold degenerate. It has been shown in \cite{HezelPRA} that the inter-$n$-manifold couplings originating from the constant and the linear term in the magnetic field, $\bm B_c$ and $\bm B_l$, are negligible. The quadratic contribution $\mu|\bm B_q(\bm R)|$ does not further constrain the parameter regime where this approximation is valid. We can therefore restrict our study to a single sub-manifold with a given principal quantum number $n$. Considering the expressions $2\braopket{\alpha'}{x_ip_j}{\alpha} =\epsilon_{ijk}\braopket{\alpha'}{L_k}{\alpha}$, we find
\begin{equation}                                                   \label{e:UHUwithHQ} 
  H \approx 
  \frac{\bm{P}^2}{2M} 
  +\bm\mu\cdot\bm B(\bm R)
 + \mathcal H_\gamma 
 +\mathcal H_Q
+H'
\end{equation}
where we named $ \mathcal H_Q:=2QZ (zL_z -xS_x-yS_y+2zS_z)$, $\bm\mu :=\frac{1}{2}\bm L_{\bm r} +\bm{S}$, and $\mathcal H_\gamma=G(xyp_z +xS_x -yS_y)$ as in Ref.~\cite{HezelPRA}. The terms  
\begin{align} 
  H'={}&\bm{S}\cdot\bm B_q(\bm r)-2Q(zXS_x+zYS_y+(xX+yY)S_z)\nonumber\\
  &+\frac{Q}{4}[( x^2y +y^3 -4yz^2)p_x      + (- xy^2-x^3+4xz^2)p_y \nonumber\\  
  & \qquad + (-y^2p_y-3x^2p_y +4z^2p_y +2xyp_x)X \nonumber\\ 
  & \qquad + (\phantom{-} x^2p_x +3y^2p_x -4z^2p_x -2xyp_y)Y ]
\end{align}
are small corrections as long as $4\frac{|Q|}{G}n^2\ll 1$. For \B{10}, \G{1} and $n=30$ this condition reads $\tilde{Q}\ll 1.5 \times 10^{-11}$. To reach geometric parameters $\tilde Q$ as large as $10^{-11}$, the coils of the Ioffe-Pritchard trap would have to be as close as $12\, \mu$m. For a macroscopic trap, the above condition is therefore always valid. In contrast to the finite-size term $\mathcal H_\gamma$, the term $\mathcal H_Q$ depends on the center of mass position. Since $L_z$ and $S_i$ are diagonal in the hydrogen basis, the latter is proportional to a dipole matrix element and hence has only off-diagonal matrix elements. Comparing its second order energetic contribution with $|QL_zZ^2|$ (which is a part of $\bm\mu\cdot\bm B$), and assuming the energetic gap of adjacent surfaces to be $B/2$, cf.\ Ref.~\cite{HezelPRA}, we find it to be negligible as soon as $ 4n^5\tilde Q\ll 1$. In macroscopic traps this restriction can only be broken with principal quantum numbers $n$ of the order of several hundreds.

Our working Hamiltonian thus reads
\begin{align}                                                  \label{e:HIP_Zconfined} 
  \mathcal H_{\text{IP}}
  =& \frac{\bm{P}^2}{2M} 
  +\bm\mu\cdot\bm B(\bm R)
  + \mathcal H_\gamma=: \frac{\bm{P}^2}{2M} +  \mathcal H_{\text{e}}.
\end{align}
In order to solve the remaining coupled Schr\"odinger equation, we adiabatically separate the relative and the center of mass dynamics by projecting Eq.~(\ref{e:HIP_Zconfined}) on the electronic eigenfunctions $\varphi_\kappa$ that parametrically depend on the center of mass coordinates:
\begin{equation}                                           \label{e:internalproblemQ}
    \mathcal H_{\text{e}} \; |\varphi_\kappa(\bm r;\bm R)\rangle 
    = E_\kappa(\bm R)\; |\varphi_\kappa(\bm r;\bm R)\rangle.
\end{equation}
We are thereby led to a set of decoupled differential equations governing the adiabatic center of mass motion within the individual three-dimensional energy surfaces $E_\kappa(\bm{R})$, i.e., the surfaces $E_\kappa(\bm{R})$ serve as potentials for the center of mass motion of the atom. The non-adiabatic (off-diagonal) coupling terms that arise within this procedure in the kinetic energy term can be neglected in our parameter regime since they are suppressed by the splitting between adjacent energy surfaces \cite{HezelPRA}.

%%%%%%%%%%%%%%%%%%%%%%%%%%%%%%%%%%%%%%%%%%%%%%%%%%%%%%%
\subsection{Electronic potential energy surfaces\label{ss:pes}}
Approximate expressions for the potential energy surfaces $E_\kappa(\bm R)$ can be found analytically when the ratio of the magnetic field gradient and the Ioffe field is small, more specifically if $n^2G/B\ll 1$. In this case, the finite size term $\mathcal H_\gamma$ is negligible compared to the contribution of $\bm\mu\cdot\bm B(\bm R)$. The latter can be diagonalized by applying the spatially dependent unitary transformation 
\begin{equation}                                                                       \label{eq:diagtrafoQ}
  U=e^{-i\alpha(L_x+S_x)} e^{-i\beta(L_y+S_y)},
\end{equation}
where $\tan\alpha={B_y}({B_x^2+B_y^2})^{-1/2}$ and $\tan\beta =-B_x/B_z$. The spatial dependence is introduced by evaluating the magnetic field components $B_i$ at the Rydberg atom's center of mass position, i.e., $B_i\equiv B_i(\bm R)$. We note that the field-free {Ha\-mil\-to\-ni\-an} $H_A$ is invariant under the transformation $U$. Moreover, 
\begin{equation}                                                        \label{eq:UmuUdaggerB}
  U\bm\mu U^\dagger \bm B= \frac{1}{2}(L_z+2S_z)|\bm B|,
\end{equation}
where $L_z$ and $S_z$ are now defined with respect to the local quantization axis. This procedure is equivalent to rotating the system into the local magnetic field direction. The adiabatic potential energy surfaces hence read
\begin{align}                                                        \label{e:EkappaQ}
  E_\kappa (\bm R)  
  \approx &(\frac{m_l}{2}+m_s) |\bm B(\bm R)|.
\end{align}
Expanding the absolute value of the magnetic field around its minimum in the trap center,
\begin{align}                                                        \label{e:EkappaQonZaxis}
  |\bm B(0,0,Z)| &\approx 2QZ^2=B(1+2\tilde QZ^2)\ , \nonumber\\
  |\bm B(X,Y,0)| &\approx B +\left(\frac{G^2}{2B} -Q\right)\rho^2 +\mathcal O(\rho ^4) \ ,
\end{align}
yields the harmonic confinement known from ground state atoms in a Ioffe-Pritchard trap ($\rho=\sqrt{X^2+Y^2}$).

In the considered limit, $n^2G/B\rightarrow 0$,  the energetically uppermost electronic adiabatic potential energy surface is the only non-degenerate one. The electronic state that corresponds to the uppermost surface is (in the rotated frame of reference) the circular one whose angular momentum projection quantum number $m_l=l=n-1$ is maximal within the given $n$-manifold. Hence, the trap frequency experienced in this surface exceeds the one of a ground state atom by a factor~$n-1$. This entails extremely large transversal trap frequencies for the external motion such that the extension of the center of mass wave function can become even smaller than the extension of the electronic cloud of the Rydberg atom \cite{HezelPRA}. Since the uppermost surface additionally suffers from the smallest non-adiabatic couplings and due to its non-degeneracy, it is best suited for a controlled confinement.

In the derivation of Eq.~(\ref{e:EkappaQ}) we neglected the finite size term $\mathcal H_\gamma$ since it only involves relative coordinates and therefore constitutes to first order solely a constant energy offset to the surfaces. Because of the coupling of the relative and the center of mass motion, however, its contribution to the electronic energy and the electronic wave function will ultimately depend on the center of mass coordinates. The resultant admixture of other hydrogenic states to the circular state entails a non-zero permanent electric dipole moment outside the trap center, which will be discussed below. The corresponding energy surface itself, on the other hand, hardly shows any deformation even for large magnetic field gradients. 
%

%%%%%%%%%%%%%%%
\subsection{Electric dipole moments\label{ss:edm}}
Let us proceed by studying the expectation value of the electric dipole moment of the Rydberg state in the uppermost adiabatic energy surface. A non-zero dipole moment arises from parity symmetry breaking terms in the Hamiltonian. The only such term in the working Hamiltonian (\ref{e:HIP_Zconfined}) is the finite-size term $\mathcal H_\gamma$, whose implicit $\bm R$-dependence will eventually entail a spatially dependent dipole moment. In the following, we pursue a perturbative treatment of $H_\gamma$ that gives rise to an explicit expression of the resulting dipole moment. To this end, the off-diagonal matrix elements of the perturbation operator $H_\gamma\sim G$ need to be much smaller than the corresponding unperturbed energy level spacings $\Delta E\sim |\bm B|$. This yields the requirement $B/G\gg n^2$ which is easily satisfied for typical Ioffe field strengths.

The perturbative treatment is detailed in Appendix~\ref{appendix:PT}. It results in the expression for the permanent electric dipole moment
\begin{align}                                         \label{e:Dgammaspacialdependencesim}
 \bm d_\gamma (\bm R)
 =&  \lambda \chi \big[ 
 -\cos\alpha\sin\beta\cos\beta   \left(\begin{array}{c}\cos\beta\\0\\\sin\beta\end{array}\right) 
 \nonumber\\
 & \phantom{ \lambda \chi \big[ }
 -\sin\alpha\cos\alpha(1+\sin^2\beta) 
 \left(\begin{array}{c}\sin\alpha\sin\beta\\\cos\alpha\\-\sin\alpha\cos\beta\end{array}\right) 
 \big] \nonumber\\
 =&
 \frac{\lambda\chi}{|\bm B|^3}
 \left(\begin{array}{c} B_x (2 B_y^2 + B_z^2)\\ -B_y (2 B_x^2 + B_z^2)  \\ (-B_x^2 + B_y^2) B_z \end{array}\right)
 \sim n^4\frac{G}{|\bm B|} \ ,
\end{align}
where $\chi:=9n^2(2n^2 -3n -\sqrt{4n^2 -10n +6} +1) /(8\Delta E)$, 
$\lambda=G/3$,
and $\Delta E\approx |\bm B|/2$ is the energetic gap between the uppermost surfaces at the trap center.  
As can be deduced from Eq.~(\ref{e:Dgammaspacialdependencesim}), the electric dipole moment is perpendicular to the local direction of the magnetic field, $\bm d_\gamma \cdot \bm B =0$. It vanishes on the $Z$-axis.

Further control of the electric dipole moment, both regarding the magnitude as well as the steric properties, can be gained by applying an additional electric field. In the following, we thus consider a modified Ioffe-Pritchard trap with an additional electric field $\bm F=(F_x,0,0)$ pointing in the $x$-direction as in Ref.~\cite{mayle:113004}. The latter can be treated perturbatively as long as $F_x \ll B/n$. The elaboration of the perturbative treatment (presented in Appendix~\ref{appendix:PT}) shows that the energetic contribution of the electric field Hamiltonian is of second order, 
\begin{equation}
 \lambda_F^2 \epsilon^{(2,F_x)}
 =  \frac{9}{4}\frac{F_x^2}{\Delta E} n^2(n-1) (\cos{\beta}^2 +\sin{\alpha}^2 \sin{\beta}^2),
\end{equation}
where $\lambda_F=|\mathbf{F}|/|\mathbf{B}|$. For vanishing $Q$ and with the approximate expression for the energetic separation between the coupling surfaces, $\Delta E \approx |\bm B|/2$, this simplifies to 
\begin{equation}                                                      
 \lambda_F^2 \epsilon^{(2,F_x)}
 \approx  \frac{9}{4} F_x^2 n^2(n-1) \frac{B^2+G^2Y^2}{B^2+G^2Y^2+G^2X^2} \ .
\end{equation}
The perturbative contribution to the uppermost surface due to an external electric field is thus positive and it is maximal on the $Z$ axis. For small atomic displacements $|\bm R|\ll B/G$, it can be considered a mere offset to the uppermost surface.

The electric field induces an electric dipole moment
\begin{equation}                                         
 \bm d_F 
 = \frac{9}{4}\frac{F_x}{\Delta E} n^2(n-1)
\frac{1}{\bm B^2}\left(\begin{array}{ccc}  B_y^2+B_z^2\\ -B_xB_y\\ -B_xB_z \end{array} \right) 
 \sim n^3 \frac{F_x}{|\bm B|} \ ,
\end{equation}
that depends on the ratio of the field strengths, as expected, and on the cubed principal quantum number. Surprisingly, however, only on the $Z$-axis it points along the direction of the generating electric field. Similar to $\bm d_\gamma$, $\bm d_F$ is in general perpendicular to the local quantization axis, which is set by the magnetic field direction, i.e., $\bm d_F \cdot \bm B = \bm d_\gamma \cdot \bm B=0$. In addition, $\bm d_\gamma$ and $\bm d_F$ are perpendicular to each other on the $Y$-axis; on the positive $X$-axis they are parallel while being anti-parallel on the negative $X$-axis. These properties only apply as long as the perturbative treatment is applicable. Our numerical results show that the dipole moment aligns with the electric field for a larger electric field strength.

%% FIGURE %%%%%%%%%%%%%%%%%%%%%%%%%%%%%%%%%%%%%%%%%%%%%%%%%%%%%%%
\begin{figure*}
   \includegraphics[width=\textwidth]{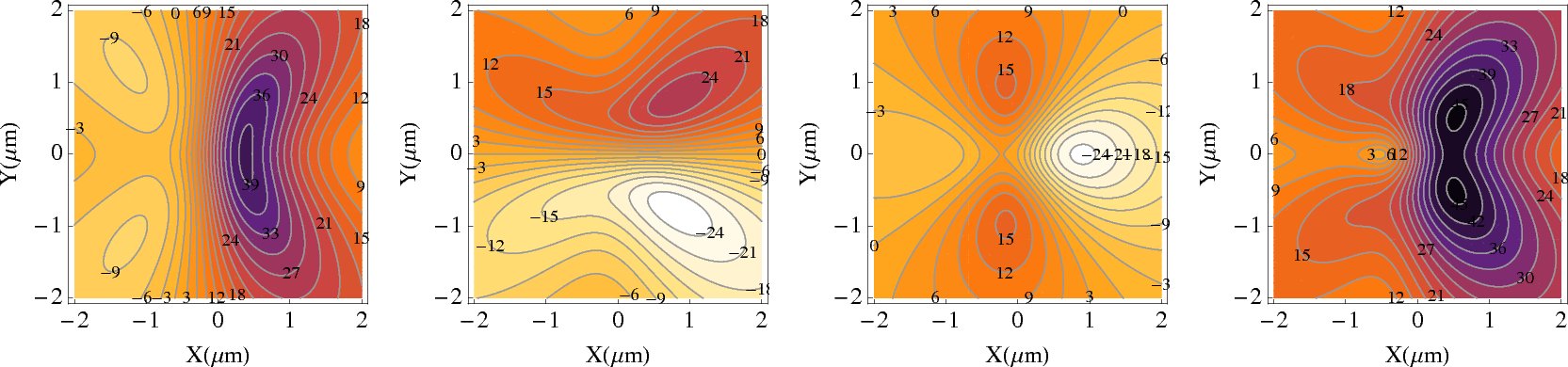}\\
\vspace{-.5cm}
\caption{                                                        \label{pic:d}
(color online) Components $d_x$, $d_y$, $d_z$ and absolute value $|\bm d|$ (from left to right) of the electric dipole moment $\bm d$ at $Z=0$.
The only symmetry that survives when a small electric field is present is the symmetry with respect to the $X$-axis, see Eq.~\ref{e:symmetriesdFgamma}. Parameters: \B{0.1}, \G{10}, \F{14}$=5.14\cdot 10^{-5}$Vcm$^{-1}$. Electric dipole moments are given in  atomic units $ea_0=2.54$ Debye.
}
\end{figure*}
%% FIGURE %%%%%%%%%%%%%%%%%%%%%%%%%%%%%%%%%%%%%%%%%%%%%%%%%%%%%%%

Analyzing the symmetry properties of the combined electric dipole moment $\bm d=\bm d_\gamma+\bm d_F$ reveals that even in case of a longitudinal confinement an approximate symmetry in $Z$ remains, $ \bm d(X,Y,Z)\approx -\bm d(X,Y,-Z)$. Much stronger than the $Z$-dependence of the electric dipole moment is its dependence on the transversal coordinates $X$ and $Y$. If no electric field is present, the relevant electric dipole moment is $\bm d_\gamma$, generated by the finite size term $\mathcal H_\gamma$. The symmetry properties of its components in the $XY$-plane read 
\begin{align}                                                        \label{e:symmetriesdgamma}
  \left(  \begin{array}{c} d_{\gamma x}\\  d_{\gamma y}\\  d_{\gamma z} \end{array} \right) ( X, Y) 
& =  \left( \begin{array}{c} \pm d_{\gamma x}\\ \mp d_{\gamma y}\\  d_{\gamma z} \end{array} \right) (\pm X,\mp Y) 
\nonumber\\
& =  \left( \begin{array}{c}-d_{\gamma x}\\ -d_{\gamma y}\\  d_{\gamma z} \end{array} \right) (-X,-Y).
\end{align}
We have furthermore $d_{\gamma,z}(X,Y) =-d_{\gamma,z}(\pm Y,\pm X)$ and $d_{\gamma,x}(X,Y) =d_{\gamma,y}(Y,X)$. The dipole moment induced by the external electric field, $\bm d_F$, exhibits different symmetries,
 \begin{align}                                                                           \label{e:symmdFplusminus}
   \left( \begin{array}{c} d_{Fx}\\  d_{Fy}\\  d_{Fz} \end{array} \right) (X, Y)
   = &  \left( \begin{array}{c} +d_{Fx}\\ -d_{Fy}\\ \pm d_{Fz}\\ \end{array} \right) ( \pm X, \mp Y) \nonumber\\
   = &  \left( \begin{array}{c} +d_{Fx}\\ +d_{Fy}\\ - d_{Fz}\end{array} \right) (-X, -Y) \ .
\end{align}
The sum of the contributions is therefore only symmetric with respect to a reflection about the $X$-axis ($Y\rightarrow-Y$):
\begin{equation}                                                        \label{e:symmetriesdFgamma}
  \bm d=
  \left( \!\! \begin{array}{c} d_{x}\\   d_{y}\\  d_{z} \end{array} \!\! \right) ( X, Y)= 
  \left( \!\! \begin{array}{c} d_{x}\\ -d_{y}\\  d_{z} \end{array} \!\! \right) ( X,-Y) \ .
\end{equation}
This can be seen in Fig.~\ref{pic:d} where the components and the absolute value of $\bm d$ are depicted. The parameters are chosen such that $\bm d_\gamma$ and $\bm d_F$ are of the same order of magnitude. Already for moderate electric fields, however, $\bm d_\gamma$ is a mere perturbation to $\bm d$ and the symmetry properties of $\bm d$ are approximately those of $\bm d_F$, cf.\ Eq.~(\ref{e:symmdFplusminus}). We note that the correct procedure   in the combined magneto-electric trap is to consider the perturbation operator consisting of the sum of the electric field term $\mathcal H_F$ and the finite size term $\mathcal H_\gamma$, rather than adding up the dipole moments $\bm d_\gamma$ and $\bm d_F$ generated by the individual contributions. However, as elucidated in Appendix~\ref{appendix:PT}, the latter approach is exact in first order. We found very good agreement of the perturbatively computed expectation values for $\bm d$ with the numerically calculated values, both for vanishing as well as for (small) finite electric field strengths.

%%%%%%%%%%%%%%%%%%%%%%%%%%%%%%%%%%%%%%%%%%%%%%%%%%%%%%%%%%%%%%%%%%%%%%%%%%%%%%%%%%%%%%%%%%%%%% 
%%%%%%%%%%%%%%%%%%%%%%%%%%%%%%%%%%%%%%%%%%%%%%%%%%%%%%%%%%%%%%%%%%%%%%%%%%%%%%%%%%%%%%%%%%%%%%
\section{Interaction-induced stabilization\label{s:iis}}

In this section, we extend our studies by considering two Rydberg atoms $A$ and $B$ that are trapped in a Ioffe-Pritchard trap and that interact via their electric dipole moments. The adiabatic Schr\"odinger equation for the two-atom center of mass wave function, $\ket{\Psi_{AB}}$, for this situation reads
\begin{multline}                                                     \label{e:cmHamiltonianwithVdd}  
\big[ T_A+T_B+V_A(\bm R_A)+V_B(\bm R_B)\\ +V_{\text{int}}(\bm R_A,\bm R_B)\big] \ket{\Psi_{AB}}=E \ket{\Psi_{AB}}\; ,
\end{multline}
where $V_{\text{int}}$ contains the interaction energy that depends on the positions of both atoms. The one-atom potential for the atoms $A$ and $B$, $V_A(\mathbf{R})=V_B(\mathbf{R})$, can be approximated for high-Ioffe-con\-fi\-gu\-ra\-tions by the analytically diagonalized term (\ref{eq:UmuUdaggerB}). The interaction potential $V_\text{int}$ will be discussed in detail in subsection \ref{s:RRI}. Because of the strong transversal confinement in the considered Ioffe-Pritchard trap, we restrict our considerations in a first step to the $Z$-axis in subsection~\ref{c:1Dstableconfiguration}. In this simplified geometry we analytically find a stable configuration of the atoms in which their distance is easily tunable without affecting neither stability nor trap frequencies. In subsection~\ref{s:3Dstableandcollaps} we extend our considerations to three dimensions and dwell on the question of stability. The last subsection is dedicated to experimental implementations suggesting different ways of realizing the system.

%%%%%%%%%%%%%%%
\subsection{Rydberg-Rydberg interaction\label{s:RRI}}
The interaction energy $V_{\text{int}}$ of two Rydberg atoms -- each modeled by a core and an electron -- can be formulated using the electric dipole moments of the individual atoms as long as the inter-atomic distance is large compared to the distance of the electrons to their respective cores. To this end, we write the Coulomb interaction between the charges of the different atoms,
\begin{multline}                                                \label{e:Coulombinteraction}
 \frac{V_{\text{int}}(\bm r_A,\bm r_B, \bm R_{AB})}{e^2/4\pi\epsilon_0} 
 =\frac{1}{|\bm R_{AB}|}
 -\frac{1}{|\bm R_{AB} -\bm r_B|}
\\
 -\frac{1}{|\bm R_{AB} +\bm r_A|}
 +\frac{1}{|\bm R_{AB} -(\bm r_B -\bm r_A)|}\ ,
\end{multline}
as a multipole expansion in the small parameter $\lambda_{\text{int}}=\langle r_{A,B}\rangle/R_{AB}$. Here, we abbreviated the vector connecting the ionic cores by $\bm R_{AB} := \bm  R_A -\bm  R_B$ and $\bm r_{A/B}$ denotes the electronic relative vectors with respect to the cores $A,B$. Due to the neutrality of the interacting constituents,  the only non-vanishing term up to third order in the expansion of $V_{\text{int}}$ is the dipole-dipole term $V_{\text{dd}}$. If we abbreviate the projections of the electronic coordinates onto the vector connecting the cores as ${r}^{P}_i :=\bm r_i\cdot\hat{\bm R}_{AB}$ and ${r}^{P}_{AB} :=(\bm r_A - \bm r_B)\cdot\hat{\bm R}_{AB} ={r}^{P}_A -{r}^{P}_B$, $\hat{\bm R}_{AB} = \bm R_{AB}/R_{AB}$, the multipole terms up to fourth order in the expansion of the interaction potential,  
\begin{equation}
  V_{\text{int}}(\bm r_A,\bm r_B, \bm R_{AB})=V_{\text{dd}} +V_{\text{dq}} +\mathcal O(\lambda_{\text{int}}^5),
\end{equation}
can be rewritten as follows:
\begin{align}                                         \label{eq:multipolepot3rd4th}
  \frac{V_{\text{dd}}R_{AB}^3}{e^2/4\pi\epsilon_0} 
  &=
  \left( \bm r_A\cdot\bm r_B -3 {r}^{P}_A {r}^{P}_B \right),  \nonumber \\
  \frac{V_{\text{dq}}R_{AB}^4}{e^2/4\pi\epsilon_0} 
  & =
  \frac{3}{2} 
  \left( r_B^2 {r}^{P}_A -r_A^2 {r}^{P}_B
    +(5 {r}^{P}_B {r}^{P}_A -2\bm r_A\cdot\bm r_B) {r}^{P}_{AB} \right) \nonumber \\
  &\approx
  \frac{3}{2} 
  \left(  (r^2 +5 {r}^{P}_B {r}^{P}_A -2\bm r_A\cdot\bm r_B) {r}^{P}_{AB} \right).
\end{align}
The last line in (\ref{eq:multipolepot3rd4th}) holds if $r_A^2\approx r_B^2$, e.g., for two circular Rydberg atoms in the same $n$-manifold. In this case the dipole-quadrupole interaction $V_{\text{dq}}$ vanishes if ${r}^{P}_{AB}$ vanishes, that is when the electric dipole moment expectation values for both atoms are identical.

%%%%%%%%%%%%%%%%%%%%%%%%%%%%%%%%%%%%%%%%%%%%%%%%%%%%%%%
If the interaction operator $V_{\text{int}}$ is treated as a perturbation to the electronic Hamiltonians of the individual Rydberg atoms, $H_A$ and $H_B$, it is favorable to represent $V_{\text{int}}$ in single-atom electronic eigenstates. We hence use the two-electron basis $\{ \ket{\varphi_i^A; \varphi_j^B} \}  \equiv  \{ \ket{ij} \}$, where $i$ and $j$ number the single-atom adiabatic electronic wave functions in the rotated frame of reference:
\begin{equation}
 (H_A+H_B) \ket{\varphi_i^A; \varphi_j^B} =(E_i+E_j) \ket{\varphi_i^A; \varphi_j^B} \ .
\end{equation}
Note that we omitted the antisymmetrization of the two electrons, which is valid in the asymptotic region we are considering where the electrons are well localized at the respective Rydberg atoms.
The leading order of $V_\text{int}$ is given by the dipole-dipole interaction operator which can be represented in the above basis as
\begin{align}                                                          \label{eq:DDIO}
 &\frac{R_{AB}^3 \braopket{i'j'}{V_{\text{dd}}}{ij}}{e^2/4\pi\epsilon_0} 
 =\braopket{i'}{\bm r}{i}\cdot\braopket{j'}{\bm r}{j} -3 \braopket{i'}{r_A^P}{i}\braopket{j'}{r_B^P}{j}\nonumber\\
 &=\bm d_{i'i}(\bm R_A)\cdot\bm d_{j'j}(\bm R_B) -3 d_{i'i}^P(\bm R_A,\bm R_{AB})d_{j'j}^P(\bm R_B,\bm R_{AB}),
\end{align}
where $d_{ij}^P:= \braopket{i}{\bm r}{j}\cdot\hat{\bm R}_{AB}$. For configurations close to the $Z$-axis, i.e., when $\hat{\bm R}_{AB} \approx (0,0,1)$, the last term can be approximately written involving the $z$-components of the electric dipole moments only, $d_{i'i}^P(\bm R_A,\bm R_{AB})d_{j'j}^P(\bm R_B,\bm R_{AB})\approx d_{i'i,z}(\bm R_A)d_{j'j,z}(\bm R_B)$.

For finite interaction between the atoms, the two-atom basis states $\ket{ij}$ are no longer eigenstates of the system. Looking at the state where both atoms are circular, $\ket{\psi_c;\psi_c}=\ket{11}$, the only non-vanishing transition dipole matrix elements are
\begin{equation}                                                    \label{e:VddonZaxiswithd13}
  \braopket{11}{V_\text{dd}}{33}
  = \frac{9}{4}\frac{n^2(n-1)}{R_{AB}^{3}} \ .
\end{equation}
where $|3\rangle$ denotes the state with $l=m_l=n-2$ in the rotated frame of reference. This coupling is small as long as
\begin{equation}
  \braopket{11}{V_{\text{dd}}}{33}\ll \delta \ \Leftrightarrow\ 
  R_{AB}\gg n \left(\frac{9}{4|\bm B|}\right)^{1/3}=R_c
\end{equation}
where $\delta\approx |\bm B|$ is the energetic separation of the surfaces. For the parameters \B{10} and $n=30$ this yields $R_{AB}\gg 1.3$~$\mu$m, which allows us to use the form (\ref{eq:DDIO}) in first order in the following, i.e., only considering diagonal elements.

%%%%%%%%%%%%%%%%%%%%%%%%%%%%%%%%%%%%%%%%%%%%%%%%%%%%%%%
\subsection{{One-dimensional stable configuration}\label{c:1Dstableconfiguration}}

A Ioffe-Pritchard trap can provide an extremely strong confinement for Rydberg atoms in the transversal, i.e., $XY$-direction \cite{HezelPRA}. We now want to take advantage of this peculiarity in order to restrict the study of the total potential $V_{\text{tot}}:=V_A+V_B+V_{\text{dd}}$ to the $Z$-axis. In Section \ref{s:collaps} we investigate the requirements on the magnetic field parameters to guarantee that this simplification is permitted. There we find that for large enough gradients $G$ this is always the case since they entail strong transversal confinement. To simplify the situation even further we impose an external electric field pointing in the $X$-direction that keeps the atoms away from each other and prevents autoionization \cite{mayle:113004}. As before, the following discussion focuses on the uppermost potential surface, emanating from the circular state.

\subsubsection*{Small oscillations of generalized coordinates} 

In order to study the one-dimensional configuration  we set the coordinates $X$ and $Y$ to zero and assume $Q$ to be non-zero and positive which generates the confining potential in the $Z$-direction. Please note that in this case ${r}^{P}_i \equiv\bm r_i\cdot\hat{\bm R}_{AB}=z_i$ in Eq.~(\ref{eq:multipolepot3rd4th}), since the atomic separation vector coincides with the $Z$-axis. We calculate the expectation value of the electric dipole moment of the single-atom eigenstate $\ket{\psi_c}$ via Eq.~(\ref{eq:DDIO}). Additionally inserting the trapping potential Eq.~(\ref{e:EkappaQonZaxis}), naming the atoms such that $Z_A>Z_B$, and omitting the constant potential offset $2nB$, the two-atom potential represented in the single-atom electronic eigenfunctions reads (again in the rotated frame of reference; ${4\pi\epsilon_0}/e^2=1$~a.u.) 
\begin{align}                                                \label{e:VtotcircZAZB}
  V_\text{tot}^{\ket{\psi_c}}
  & (Z_A, Z_B) 
  :=  \braopket{\psi_c,\psi_c}{V_\text{tot}(Z_A, Z_B)}{\psi_c,\psi_c}\nonumber \\
  =&2nQ(Z_A^2+Z_B^2)\nonumber\\
  &+\frac{81 F_x^2n^4(n-1)^2}{4(B+2QZ_A^2)(|Z_A-Z_B|)^3(B+2QZ_B^2)} .
\end{align}
Utilizing generalized coordinates for the distance of the atoms and for their center of mass, $Z_D=Z_A-Z_B>0$ and $Z_S=({Z_A+Z_B})/{2}$, respectively, the total potential (\ref{e:VtotcircZAZB}) translates to 
\begin{equation}                                                \label{e:approxVtotcirconZaxis}   
V_\text{tot}^{\ket{\psi_c}}(Z_D, Z_S)
\approx \frac{81}{4}n^4(n-1)^2 \frac{F_x^2}{B^2} \frac{1}{Z_D^{3}} +nQ  \left(Z_D^2+ 4Z_S^2\right) .
\end{equation}
Here we approximated $B+2QZ_{A,B}^2 \approx B$, which is valid as long as
\begin{equation}                                \label{eq:approxAbsZSpZDhllsqrtBo2Q} 
|Z_S|+Z_D/2  \ll \sqrt{B/(2Q)} \ .
\end{equation}

The first term in Eq.~(\ref{e:approxVtotcirconZaxis}) is the approximate version of the dipole-dipole interaction operator. It only depends on the distance of the atoms. Higher order terms originate in the quadratic $Z$-dependence of the interacting electric dipole moments. They become significant only for very large $Z$ or exceptionally strong parameters $Q$ reachable on atoms chips. The coordinate for the center of mass of both atoms, $Z_S$, appears as the quadratic shift $4nQZ_S^2$. An equilibrium configuration of the atoms is therefore bound to be symmetric around the origin, i.e., $Z_S=0$. Minimizing the energy of the two-atom potential within this approximation, we find the equilibrium position at
\begin{align}                                                    \label{eq:approxZDmin} 
  Z_{S,\text{min}} &= 0 \ , \nonumber\\
  Z_{D,\text{min}} &= 3\sqrt[5]{\frac{F_x^2 (n-1)^2 n^3}{8B^2Q}}
  \approx \frac{3}{2^{3/5}} n \sqrt[5]{\frac{F_x^2 }{B^2Q}} \ .
\end{align}
The expression for the equilibrium distance, $Z_{D,\text{min}}$, can hence be readily controlled by the electric field strength $F_x$. The condition of validity of our approximation (\ref{eq:approxAbsZSpZDhllsqrtBo2Q}) at the equilibrium position (\ref{eq:approxZDmin}) reads
\begin{equation}                                     \label{eq:restrDatZDmin} 
  c_D:= \frac{3}{2}\sqrt[5]{F_x^2 (n-1)^2 n^3}\sqrt[10]{\frac{Q^3}{2B^9}}
  \approx \frac{3n}{2} \sqrt[10]{\frac{F_x^4Q^3}{2B^9}} \ll 1 \ .
\end{equation}
See Tab.~\ref{t:cD} for explicit values.
\begin{table}
  \caption{                                                                    \label{t:cD}
    Explicit values for $c_D$, Eq.~(\ref{eq:restrDatZDmin}), which measures the quality of the approximation (\ref{eq:approxAbsZSpZDhllsqrtBo2Q}) at the equilibrium position (\ref{eq:approxZDmin}). The values are computed for the geometry parameter $\tilde Q=6\times 10^{-16}$ which is around the highest values reachable with macroscopic Ioffe-Pritchard traps \cite{Moore}. The restriction $c_D\ll 1$ is violated only for very low Ioffe fields paired with electric fields that would ionize the Rydberg atoms. ($10^{-11}$~a.u.$= 5.142\,206\,32$~V/cm).  
  }
\begin{ruledtabular}
 \begin{tabular}{lccc}
   \quad F&$\ 10^{-12}$ {a.u.}&$\ 10^{-11}$ {a.u.}&$\ 10^{-10}$ {a.u.}\\
   \hline
   \B{0.1}&  0.0297 & 0.0747 & 0.1877 \\
   \B{1}&    0.0075 & 0.0188 & 0.0471 \\
   \B{10}&   0.0019 & 0.0047 & 0.0118 \\
   \B{100}&  0.0005 & 0.0012 & 0.0030 \\
 \end{tabular}
\end{ruledtabular}
\end{table}

Since we are interested in the motion of the system around a stable equilibrium configuration, we expand the potential in a Taylor series around that equilibrium and solve the corresponding classical eigenvalue problem. The resulting frequency for the center of mass and relative motion read in the harmonic approximation
\begin{equation}                                 \label{e:Findependenttrapfrequencies}
  \omega_D^2 = \frac{20nQ}{2M},\quad \omega_S^2= \frac{4nQ}{2M},
\end{equation}
where 2M is the total mass of the system.
It is worth noting that within the approximation (\ref{eq:approxAbsZSpZDhllsqrtBo2Q}) \emph{the eigenfrequencies are independent of the electric field strength $F_x$.} They indirectly depend on the Ioffe field strength since \hbox{$Q=B\tilde Q$}.

%%% FIG %%%%%%%%%%%%%%%%%%%%%%%%%%%%%%%%%%%%%%%%%%%%
\begin{figure}
\includegraphics[width=\columnwidth]{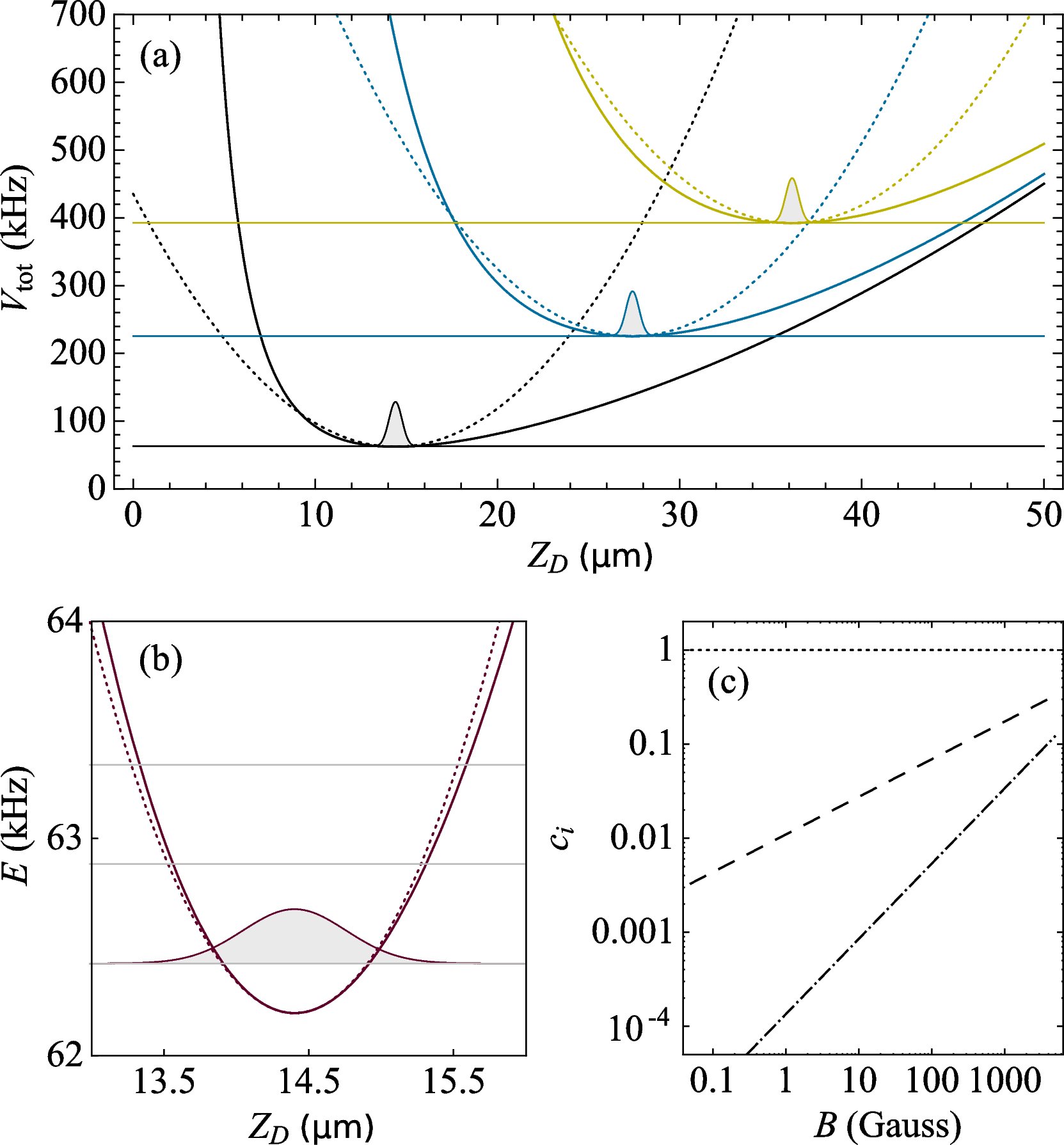}
 \caption{                                                \label{fig:EalongZDdiffFPRL2} 
   (color online) (a) Two-atom potential energy, Eq.~(\ref{e:VtotcircZAZB}) (solid), the harmonic approximation (dotted) and the 1D ground state energy along $Z_D$ for $Z_S=0$ and different electric field strengths $F$. Parameters: \B{10}, $Q=6\times 10^{-16}B$, $n=30$, $F=0.0514$~V/cm (black), $0.257$~V/cm (blue/gray), $0.514$~V/cm (yellow/light gray). (b, c) Quality of the harmonic approximation for the first set of parameters. Subfigure (b) depicts the harmonic approximation (dotted) of the two-atom potential $V_\text{tot,circ}$ (solid) in a region around the local minimum (filled: Gaussian). The double-logarithmic plot in subfigure (c) shows the quadratic (dotted), cubic (dashed) and the quartic (dot-dashed) coefficient of the expansion of the potential around the equilibrium position as a function of the Ioffe field strength.
}
\end{figure}

\subsubsection*{Tuning the distance of the atoms}
The fact that the equilibrium distance $Z_{D,\text{min}}$ of the atoms, Eq.~(\ref{eq:approxZDmin}), can be increased without changing the trap frequency by just increasing the electric field strength $F$ is depicted in Fig.~\ref{fig:EalongZDdiffFPRL2}(a). The two-atom potential and its harmonic expansion around the equilibrium position are drawn for different electric field strengths. In subfigure \ref{fig:EalongZDdiffFPRL2}(b) a magnified view of the minimum is provided, demonstrating the validity of the harmonic approximation. The figure shows that we can safely assume the \cm~ground state to be a Gaussian with the corresponding trap frequency $\omega_D$. Subfigure ~\ref{fig:EalongZDdiffFPRL2}(c) in addition shows expansion coefficients of the potential around the equilibrium position (computed without assuming the approximation (\ref{eq:restrDatZDmin}) to be valid). It is evident from the plot that the harmonic description of the potential around the equilibrium position is a good approximation for small Ioffe field strengths. Therefore, not only the trap frequency but also the center of mass ground state remains the same for different values of $Z_D$. 
To give a numerical example, the variation of the trap frequency $\omega_{D}$ does not exceed $10^{-3}$ for a Ioffe field of 1 Gauss as long as the electric field strength is smaller than $F\approx 2.3\ $V/cm.

%%%%%%%%%%%%%%%%%%%%%%%%%%%%%%%%%%%%%%%%%%%%%%%%%%%%%%%
\subsubsection*{Quadrupole-quadrupole repulsion}
In this subsection we study the influence of higher order multipole interactions and answer the question in which situations they can change the behavior of the system. We do this on the $Z$-axis for vanishing electric field and consider -- as throughout the present work -- the circular Rydberg state.

%% F=0
When no electric field is present, circular Rydberg atoms located on the $Z$-axis of a Ioffe-Pritchard trap do not exhibit a permanent electric dipole moment. They are hence neither subject to dipole-dipole interaction nor to dipole-quadrupole interaction. Since the circular electronic wave function is not spherically symmetric, they feature a quadrupole moment, however. This is a first-order effect and the repulsive quadrupole-quadrupole interaction 
is hence the leading order of the interaction potential. For the special configuration we are considering, it can be calculated using the simplified expression \cite{Dalgarno1966}
\begin{multline}                                      \label{e:QQWWonZaxis}
  V_{\text{qq}}
=\frac{3}{4Z_D^5} \big[
  r_A^2r_B^2-5(z_A^2r_B^2+r_A^2z_B^2)-15z_A^2z_B^2
\\  +2(x_Ax_B+y_Ay_B-4z_Az_B)^2  \big] \ .
\end{multline}
With the matrix elements 
$\braopket{\psi_c}{r^2}{\psi_c}=\frac{1}{4}n^2(n+1)(n+\frac{1}{2})$,
$ \braopket{\psi_c}{x^2}{\psi_c}=\braopket{\psi_c}{y^2}{\psi_c}=\frac{1}{2}n^3(n+1)$, and 
$\braopket{\psi_c}{z^2}{\psi_c}=\frac{1}{2}n^2(n+1)$,
we find 
\begin{equation}                                                        \label{e:Vqqonzaxis}
  V_{\text{qq}}=\frac{1}{Z_D^5}\frac{3}{2}n^4(n+1)^2(n+\frac{1}{2})^2
  \approx  \frac{1}{Z_D^5}\frac{3}{2}n^8 \ .
\end{equation}
For low enough \cm~kinetic energy, the repulsion of the atoms due to the quadrupole-quadrupole interaction could in principle stabilize Rydberg atoms on the $Z$-axis against auto-ionization. We must not forget, however, that the van-der-Waals coupling as a second order contribution to the multipole interaction can be of similar strength for low enough distances in the considered parameter regime. 

The situation changes completely when an electric field is applied. The induced dipole moments scale linearly with the field strength and the dipole-dipole interaction then depends quadratically on $F$. From the first term in Eq.~(\ref{e:approxVtotcirconZaxis}) its magnitude can be estimated to be 
\begin{equation}
  V_{\text{dd},\text{$Z$-axis}}\approx \frac{1}{Z_D^3}\left( \frac{9}{2}n^3\frac{F}{B}\right)^2.
\end{equation}
The dipole-quadrupole interaction happens to be zero on the $Z$-axis even for finite electric field strength. In order for the dipole-dipole interaction, $|V_{\text{dd}}|\sim |\bm d|^2/Z_D^3$, to dominate the quadrupole-quadrupole interaction, $|V_{\text{qq}}|\sim \frac{3}{2}n^8/Z_D^5$, the condition
\begin{equation}                                           \label{e:condDDWWQQWW}
  \Big|\frac{V_{\text{dd}}}{V_{\text{qq}}}\Big|
  =\frac{27}{2}\left(\frac{Z_DF}{nB}\right)^2 \gg 1
  \Leftrightarrow
  Z_D\gg \frac{1}{3}\sqrt{\frac{2}{3}} n\frac{B}{F}
\end{equation}
must be fulfilled. For a Ioffe field strength \B{10} and an electric field strength as low as \F{12} (and $n=30$) this reads $Z_D\gg 35000$~\au$=1.7\,\mu$m. Increasing the electric field strength to $F=2\times 10^{-11}$\,a.u.\ already yields $Z_D\gg 1700$~\au$=90\;$nm. For the examples above it is therefore legitimate to neglect the quadrupole-quadrupole interaction. Around the equilibrium configuration $Z_{D,\text{min}}$  of the atoms, Eq.~(\ref{eq:approxZDmin}), condition (\ref{e:condDDWWQQWW}) is even easier to fulfill.

%%% FIG %%%%%%%%%%%%%%%%%%%%%%%%%%%%%%%%%%%%%%%%%%%%
\begin{figure*}
  \includegraphics[width=.88\textwidth]{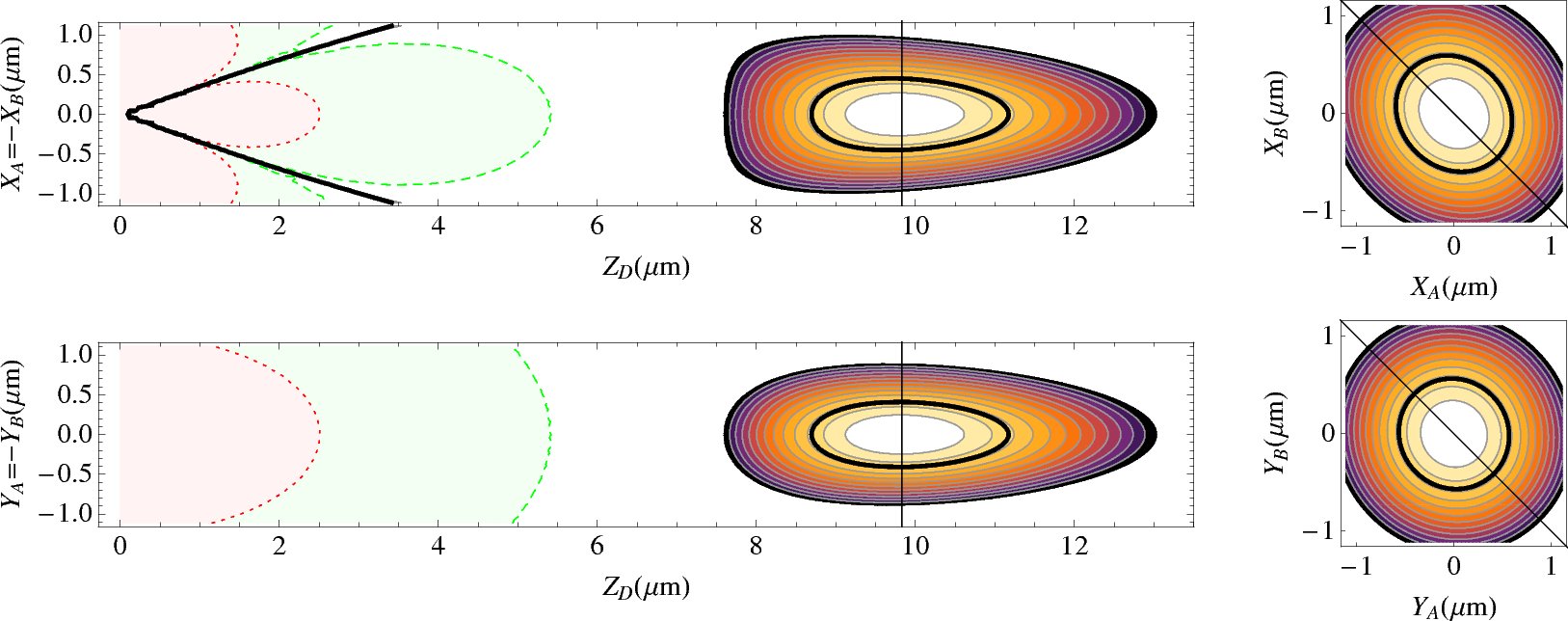}
  \caption{ \label{f:secaroundlocmin2} 
    (color online) Sections through the six-dimensional two-atom potential around the local minimum. The $Z_D$-coordinate of the minimum is indicated by the vertical lines. The thick black contours are plotted at the energy of the two-atom harmonic center of mass ground state which corresponds to half the sum of all six trap frequencies, $E_{0}=26\;$kHz. The contour plots are clipped at energies $125\;$kHz higher than the minimal energy configuration. The dashed and dotted lines in the plots on the left hand side indicate the quality of the single-atom-surfaces-approximation introduced in Sec.~\ref{s:RRI}. They are drawn where the ratio of dipole-dipole interaction energy and energetic distance of adjacent single-atom electronic surfaces, $E_\text{dd}/\Delta E$, equals $0.1$ (dotted) and $0.01$ (dashed). Parameters: \B{30}, \G{10}, $Q=6\times 10^{-16}B$, $F=2\times10^{-11}\;$a.u.$=10.28\;$V/m, $n=30$.
 }
\end{figure*}
%%% FIG %%%%%%%%%%%%%%%%%%%%%%%%%%%%%%%%%%%%%%%%%%%%

%%%%%%%%%%%%%%% 
\subsection{Three-dimensional stable configuration and collapse\label{s:3Dstableandcollaps}}

Very strong transversal confinement leads to the one-dimensional situation discussed in the preceding Chapter~\ref{c:1Dstableconfiguration}. Since the magnetic field gradient $G$ only influences the transversal but not the longitudinal confinement, the respective trap frequencies can be altered independently. If the transversal confinement is decreased and/or if the longitudinal confinement is increased the local minimum of the potential mentioned above turns into a saddle point: The tendency of the dipole-dipole interaction to force the atoms to step out of the $Z$-axis wins against the confining nature of the transversal magnetic field gradient. The atoms then attract each other and most probably eventually ionize. These statements are substantiated and refined in the following. 

The two-atom interaction potential exhibits an approximate longitudinal symmetry as can be justified on account of the smallness of $\tilde Q=Q/B$. More precisely, the conditions $2\tilde QZ^2\ll 1$ [Eq.~(\ref{eq:approxAbsZSpZDhllsqrtBo2Q})] and $4\tilde Q|Z|\ll G/B$ have to be met. Then, the dipole-dipole interaction $V_{\text{dd}}$ between two atoms in the circular state exhibits a dependency on $Z_S$ that is negligibly small. The confinement due to the Ioffe-Pritchard magnetic field configuration, on the other hand, is harmonic around the origin in all directions. Within the range of validity of the above approximations, we can thus conclude that minimizing the total energy of the two atoms always leads to a symmetric configuration, $Z_S=0$. We therefore set $Z_S=0$ for the following analysis. 

In order to characterize the six-dimensional adiabatic two-atom potential around the local stable minimum we first keep a strong transversal confinement which is quantified by the trap frequencies in $X$- and $Y$-direction. For a single atom in a Ioffe-Pritchard trap and for finite $Q$ they read
  \begin{align}
    \omega^2_X = \frac{2n}{M}\frac{G^2 -2Q(B +2GZ)}{B +2QZ^2}
    \ , \nonumber\\
    \omega^2_Y = \frac{2n}{M}\frac{G^2 -2Q(B -2GZ)}{B +2QZ^2} \ .
  \end{align}
Corrections due to the dipole-dipole interaction energy when two atoms are considered are proportional to $\frac{n^3}{M}\frac{F}{B}Z_D^{-5/2}$ for a dominating Ioffe field. 

In what follows, let us investigate the exemplary parameter set \B{30}, \G{10}, $Q=6\times 10^{-16}B$, $F=2\times 10^{-11}$\,a.u., and $n=30$. All the inequalities, that have been formulated up to now in order to measure the quality of the applied approximations, hold with a confidence factor of at least $10^2$. The only exception is the quality of the perturbative approach to determine the electric dipole moment. The corresponding requirement explicitly reads $nF_x/B=0.047\ll 1$ for the chosen parameters which is still satisfactory. 

Diagonalizing the Hesse matrix $[{\partial^2 V_\text{tot}}/({\partial R_i \partial R_j})]$ at the local minimum position of the total, i.e., two-atom potential and extracting the trap frequencies along the principal axes from the eigenvalues and its eigenvectors, respectively, yields
\begin{align}                                     \label{e:stableconfexamplefrequencies}
  11&.1 \text{kHz},\quad  (Y_A=Y_B),\nonumber\\
  11&.1 \text{kHz},\quad  (X_A=X_B),\nonumber\\
  10&.7 \text{kHz},\quad  (Y_A=-Y_B),\nonumber\\
  9&.7  \text{kHz},\quad  (X_A=-X_B),\nonumber\\
  5&.0  \text{kHz},\quad  (Z_D), \nonumber\\
  4&.5  \text{kHz},\quad  (Z_S),
\end{align}
where the equations in brackets define the directions of the principal axis. It is convenient to introduce appropriate generalized coordinates for all three spatial dimensions, namely, $\bm{R}_D=\bm{R}_A-\bm{R}_B$ and $\bm{R}_S=\frac{1}{2}(\bm{R}_A+\bm{R}_B)$. Sections of the total potential around its minimum are shown in Fig.~\ref{f:secaroundlocmin2}.

The trap frequencies in (\ref{e:stableconfexamplefrequencies}) have to be compared to the radiative lifetime of the interacting atoms. The field-free lifetime of the electronic state corresponding to the uppermost surface is $\tau(n,n-1)\approx \frac{3}{2c^2}\left(\frac{n}{\alpha}\right)^5=2.3\,$ms \cite{Jentschura2005}. Even the reduction of the lifetime due to admixtures to the pure circular state originating from the finite size term and due to the coupling to the electric field leaves it close to the field-free value \cite{mayle:113004}. We hence expect more than 10 oscillations of the atomic motion to be observable within the lifetime of the Rydberg state. Since $\omega^2\sim nQ =nB\tilde Q$ the trap frequencies $\omega$ can obviously be increased by the choice of a larger principal quantum number $n$, with higher Ioffe field strengths $B$ and by allowing for a stronger longitudinal confinement. The latter can be achieved by shrinking the trap onto an atom chip \cite{folman}. We emphasize that the stable two-atom configuration is not immediately lost when one of the atoms decays to the circular state of the adjacent $n$-manifold since the electronic properties of that state are very similar and, therefore, so are the electric dipole moment and the adiabatic surface.

\subsubsection*{Loss of confinement and collapse\label{s:collaps}}

We now study mechanisms that endanger the stability of the equilibrium position on the $Z$-axis when the transversal confinement is relaxed. We identify two situations in which this happens. One of them is the loss of the confining property of the Ioffe-Pritchard field configuration for a single atom for large ratios $B/G$. Furthermore, the stability of the equilibrium configuration is also lost as soon as the transversal confinement becomes smaller than the transversal anti-confinement due to the dipole-dipole interaction.

Regarding the first mechanism, the transversal magnetic confinement for a single atom is only guaranteed as long as the respective curvature is positive. This yields the restriction
\begin{equation}                                       \label{e:condmagnconf}
  \frac{G^2}{B^2}>2\tilde Q\;(1+4\frac{G}{B}|Z|) \ .
\end{equation}
In order for this condition to be broken at the origin of the trap, the ratio $G/B$ must be extremely small since the highest reachable values for the geometry parameter $\tilde Q$ in macroscopic Ioffe-Pritchard traps are around $10^{-15}$ (we use $\tilde Q=6\times 10^{-16}$ for all presented examples \cite{Moore}). For large enough displacements in $Z$-direction, however, the condition can always be broken. To give a sense of the numbers, we insert the exemplary parameter set \B{10}, \G{2} and  $\tilde Q=6\times 10^{-16}$ to find that the displacement $|Z|$ must be as large $2\times 10^{7}\;$a.u.~$=1\;$mm to break the condition (\ref{e:condmagnconf}). We do not consider such large atomic distances from the trap center for any example.

\begin{figure} % Dipoles 
\includegraphics[height=.135\textheight]{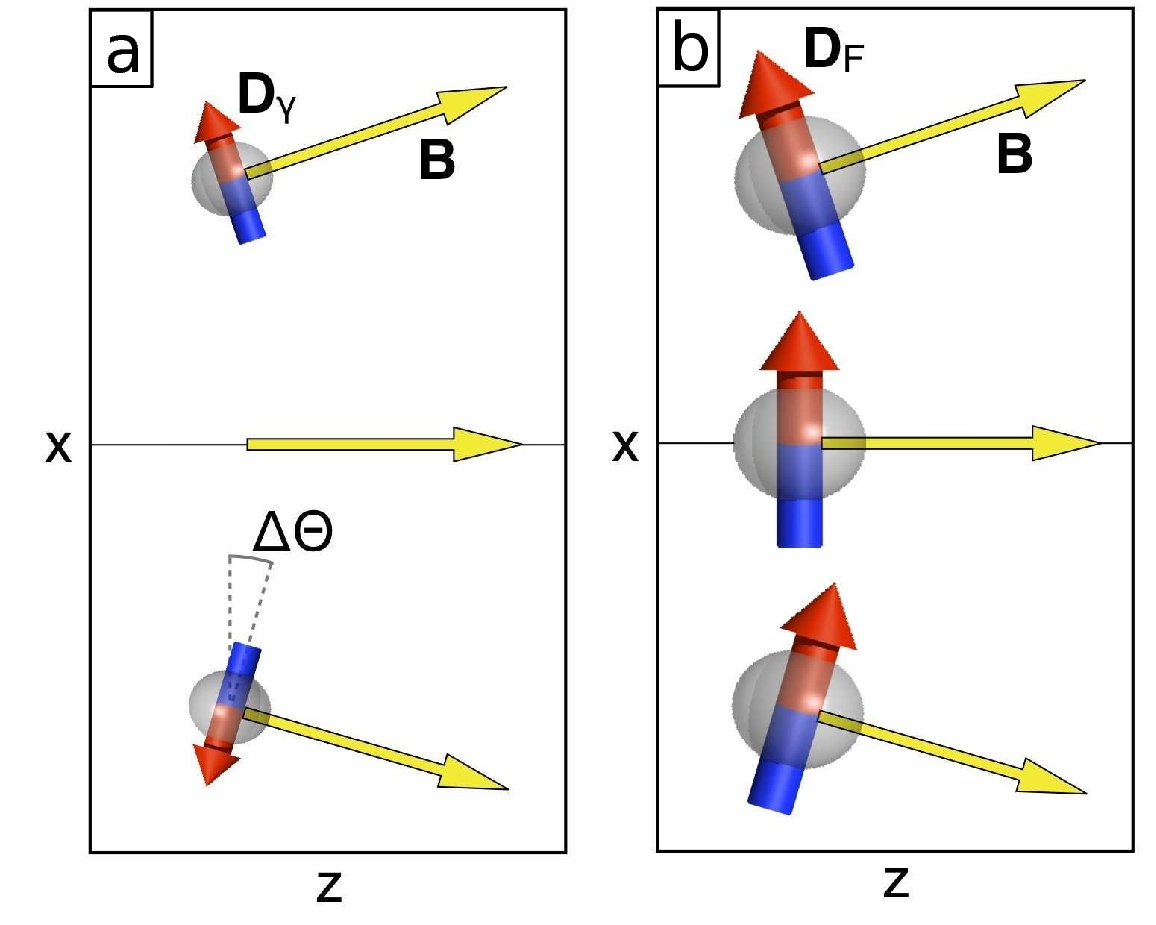} \nolinebreak
\includegraphics[height=.135\textheight]{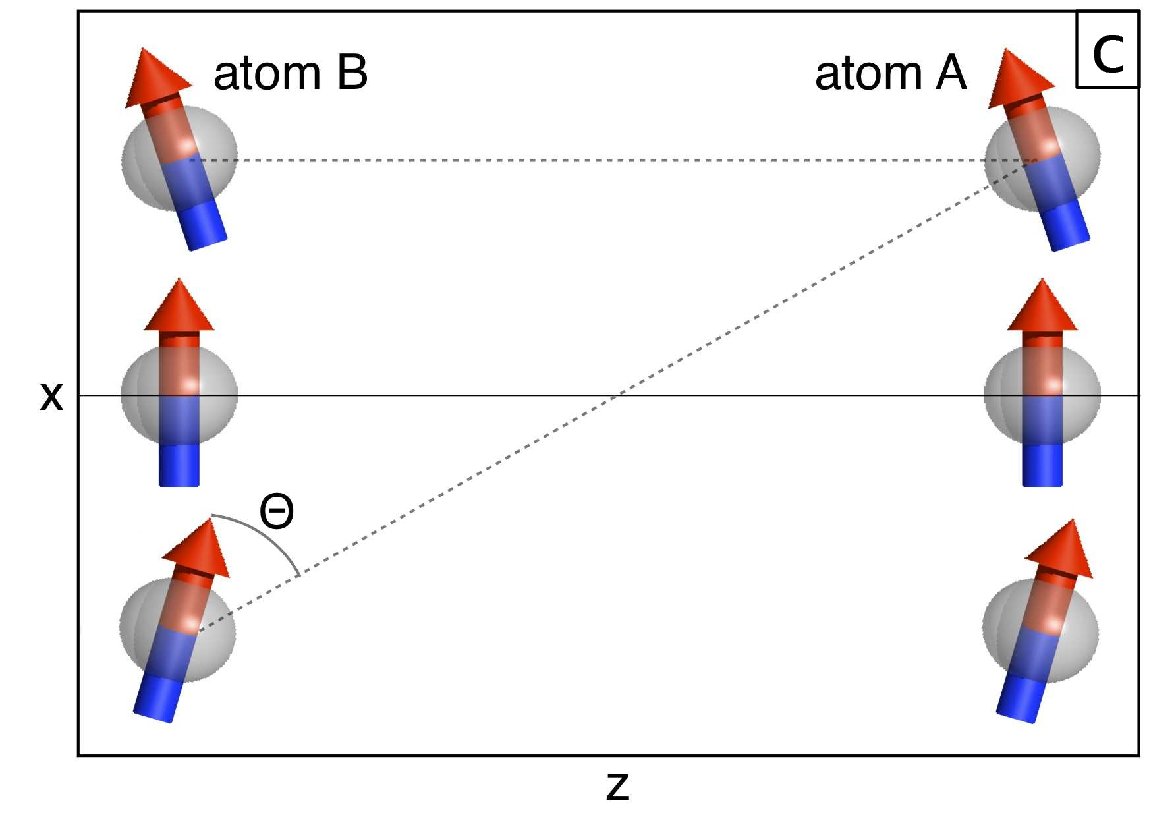}
\caption{                                                \label{fig:dipoles}
   (color online) Spatial dependence of the direction of the electric dipole moments $\bm d_\gamma$ and $\bm d_F$ (for $Q=0$). (a) The electric dipole moment $\bm d_\gamma$ that originates from the finite size of the Rydberg atom. It is perpendicular to the local magnetic magnetic field direction (yellow arrows) and it vanishes on the $Z$-axis. On the positive (negative) $X$-axis it is parallel (anti-parallel) to the electric dipole moment $\bm d_F$ that is induced by the electric field and depicted in the subfigure (b). (c) Orientation of $\bm d_F$ with respect to the vector $\bm R_{AB}$ connecting the interacting atoms $A$ and $B$ when they are displaced in the same $X$-direction, $X_S\ne 0$ (solid line), or in different $X$-directions, $X_D\ne 0$, (dashed lines).
}
\end{figure}

The second reason for the loss of the stable equilibrium configuration on the $Z$-axis is the dipole-dipole interaction between the two atoms. Besides being the interaction of longest range between neutral atoms, the major property of the dipole-dipole interaction is its anisotropic character. This comes into play when the atoms can step out of the $Z$-axis and the angles between the electric dipole moments and the connecting vector change (Fig.~\ref{fig:dipoles}). 
%%% FIG %%%%%%%%%%%%%%%%%%%%%%%%%%%%%%%%%%%%%%%%%%%%
%%% FIG %%%%%%%%%%%%%%%%%%%%%%%%%%%%%%%%%%%%%%%%%%%%

% If these angles are the same for both atoms, i.e., when the dipole moments are parallel, and assuming the strengths $d$ of the electric dipole moments to be the same, the expression for the dipole-dipole interaction operator simplifies to 
% \begin{equation}                                      \label{e:DDWWpolarized}
%   V_{\text{dd}}=  \frac{d^2}{R_{AB}^3}(1-3\cos^2{\theta}) \ ,
% \end{equation}
% where $\theta$ is the angle between the direction of the moments and the direction of the connecting vector $\bm R_{AB}$, $0<\theta <\pi /2$. The dipole-dipole interaction has then the angular symmetry of the Legendre polynomial of second order $P_2(\cos{\theta})$, i.e., d-wave \cite{Lahaye2009}. The expression in brackets in (\ref{e:DDWWpolarized}) yields one for parallel dipole moments next to each other ($\theta =\pi /2$). It becomes smaller when $\theta$ is reduced and eventually becomes negative for angles smaller than the so called magic angle $\theta^*=\arccos{(3^{-1/2})}\approx 55^{\circ}$. For the head-to-tail configuration, $\theta=0$, it yields $-2$.

\emph{Polarized case:} In order to simplify our considerations let us first assume that the external electric field $\bm F=(F_x,0,0)$ fully polarizes the atoms. This happens for relatively large electric field strengths as discussed in App.~\ref{c:electricdipolemoment_finitefield}. Then all dipole moments point in the $X$-direction. For the sake of clarity we now additionally assume that the magnitude of the dipole moments does not depend on the position of the atom in the trap. Then a displacement of both atoms $A$ and $B$ from the $Z$-axis in the same direction, $X_S\ne 0$ and/or $Y_S\ne 0$, does not change the interaction energy since the angle $\theta$ between the two dipoles does not change. The angle also stays the same for a displacement of the atoms in opposite $Y$-directions, $Y_D\ne 0$. Here the interaction energy decreases only slightly due to the increase of the distance $R_{AB}$ of the atoms. A displacement of the atoms in opposite $X$-directions, $X_D\ne 0$, however, changes $\theta$ and thereby decreases the interaction strength considerably. For a decreasing transversal confinement we therefore expect the stable configuration to collapse by a displacement of the atoms in opposite $X$-directions in the polarized case. 

\emph{Tilted moments:} The reasoning above is based on the assumption that the atoms are polarized, meaning that their dipole moments point in the direction of the electric field independent of the position of the atom. This is an oversimplification in case of Rydberg atoms in a strongly confining magnetic field configuration. In Section~\ref{ss:edm} it has been shown that the application of a moderate electric field induces a dipole moment $\bm d_F$ that is perpendicular to the local direction of the magnetic field. In the $Y$-$Z$-plane $\bm d_F$ points in the $X$-direction (we neglect here marginal dependences on $Q$). For finite $X$, by contrast, it has a finite $Z$-component, see App.~\ref{c:electricdipolemoment_finitefield}. We thus concentrate once more on displacements in $X$ and set $Y_S=X_D=0$ in the following. The non-polarized case is illustrated in Fig.~\ref{fig:dipoles}.

In case of a displacement of the atoms in opposite directions, $X_D>0$, the dipole moments of the two atoms include different angles with their connecting axis. They differ from the angles in the fully polarized case, $\theta_P$,
by the additional tilt due to the local magnetic field direction, $\pm\Delta\theta$. The interaction energy no longer depends on $1-\cos^2(\theta_P)$ but on $1-\cos(\theta_P+\Delta\theta)\cos(\theta_P-\Delta\theta)$ which is smaller than the former for $0\le\theta_P\pm \Delta\theta\le\pi /2$. For electric dipole moments perpendicular to the local magnetic field axis, as considered here, the reduction of the dipole-dipole interaction energy for displacements in $X_D$ is therefore smaller than in the fully polarized case discussed above. The curvature in that direction is thus expected to be still positive for shallower transversal confinements.

Symmetric displacements of the atoms in the same $X$ direction, $X_S\ne 0$, do not change the dipole-dipole interaction energy in the fully polarized case as stated above. For tilted dipole moments, however, the energy is reduced 
since the moments are no longer perpendicular to their connecting axis. This can be seen from the illustration in Fig.~\ref{fig:dipoles}. The additional angle due to the orientation of the moments perpendicular to the local magnetic field axis (for $Q=0$) explicitly reads 
\begin{equation}
  \Delta\theta=\arctan\left(\frac{G}{B}X\right) \ .
\end{equation}
The effects due to $\Delta\theta$ described above are therefore weak for typical parameter sets since for typical parameters the ratio $G/B$ is small. A strong effect on the dipole-dipole interaction energy is expected for large ratios $G/B$. In this case, however, the transversal confinement is strong and the curvature in $X$-direction is positive on the $Z$-axis nonetheless.

%%% FIG %%%%%%%%%%%%%%%%%%%%%%%%%%%%%%%%%%%%%%%%%%%%
% Collapse with reduction of G %%%%%%%%%%%%%%%%%%%%%%%%%%%%%%%%%%%%%%%%%%%%%%%%%%
\begin{figure*} 
\includegraphics[width=.7\textwidth]{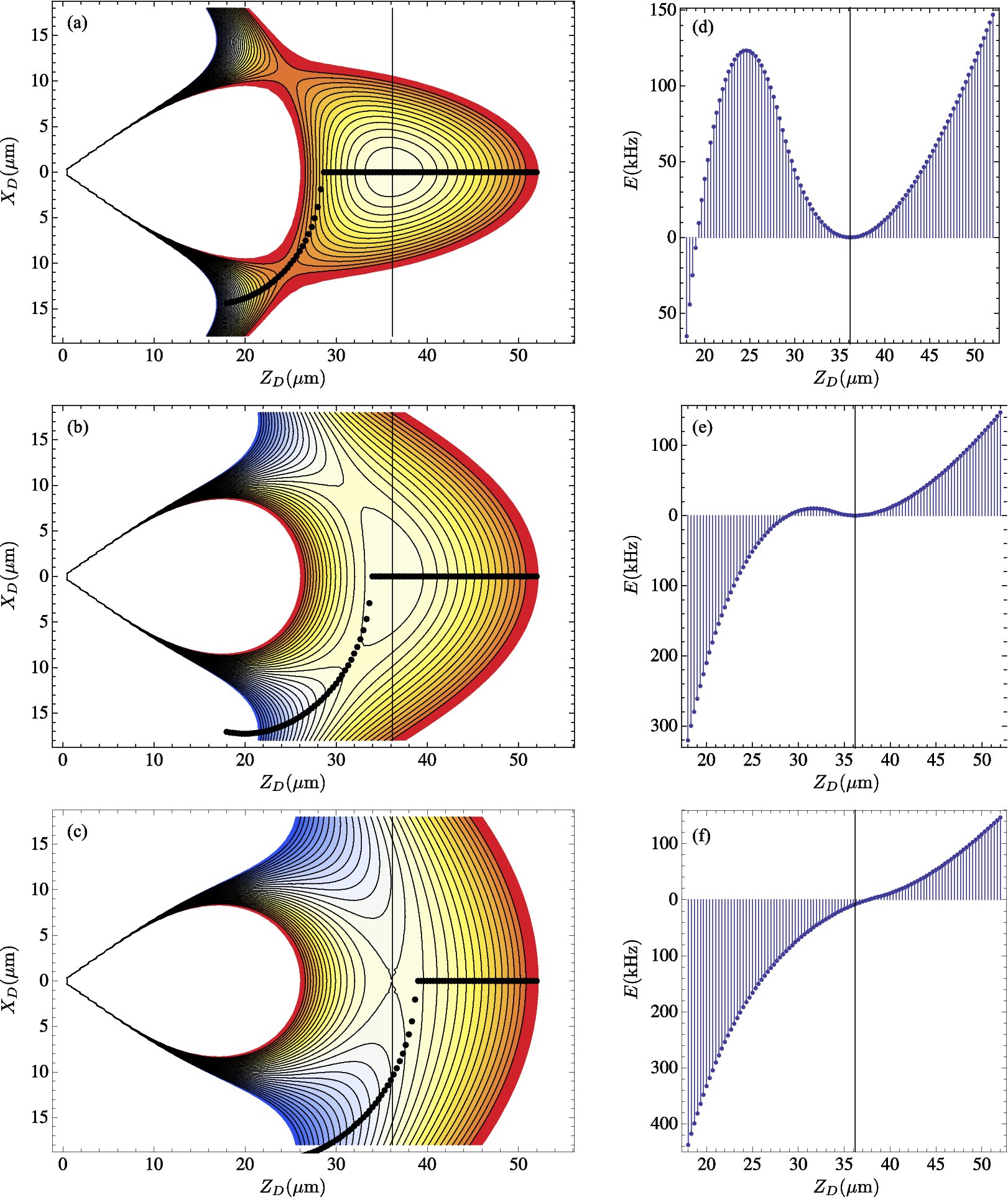}
  \caption{                                        \label{fig:implosionGmax}
     (color online) Loss of the local minimum for decreasing magnetic field gradient $G$ of the Ioffe-Pritchard trap. (a-c) Two-dimensional sections of the six-dimensional two-atom potential through the $X_D$-$Z_D$ plane, $X_S=Y_S=Y_D=Z_S=0$. 
% Red colors symbolize high energies and blue colors symbolize low energies. 
     The plot range of all three contour plots is $\pm 150$~kHz. (d-f) Minimal energy of the two-atom potential against $Z_D$. Each point is computed minimizing the total potential for fixed $Z_D$ with the other \cm~coordinates as parameters. The $X_D$ positions are shown as black dots in the sections on the left. Parameters: \B{10}, $Q=6\times 10^{-16}B$, $F=10^{-10}$\,a.u., $n=30$, \G{3} [first row, (a,d)], \G{2} [second row, (b,e)], \G{1.5} [third row, (c,f)].}
\end{figure*}

In order to verify the predictions above on how the stability of the equilibrium configuration of the atoms gets lost, and the predictions for the configuration the atoms take when they step out of the $Z$-axis, we minimize the total two-atom potential for the coordinates $X_S$, $X_D$, $Y_S$ and $Y_D$ for fixed symmetric displacements of the atoms in $Z$, $Z_D>0$ and $Z_S=0$. As discussed above we expect the atoms to align on the $Z$-axis as long as the transversal confinement dominates the interaction. When the longitudinal confinement increases and/or the transversal confinement decreases, the atoms are expected to step out of the $Z$-axis in different $X$-directions, i.e., $X_D\ne 0$, $X_S=Y_D=Y_S=0$, since the negative transversal gradient at the equilibrium position due to the dipole-dipole interaction is largest in the $X_D$-direction. This change of the atoms' configuration is depicted in Fig.~\ref{fig:implosionGmax} where we show the computed $X_D$ positions that yield minimal energy for fixed distances $Z_D$ as black dots into the two-dimensional section through the two-atom potential for different parameter sets (left plots). 
The series of plots in Fig.~\ref{fig:implosionGmax} shows the loss of the local minimum position for relaxing transversal confinement due to decreasing field gradients $G$ (from top to bottom). The vertical lines indicate the position of the equilibrium configuration on the $Z$-axis. The bar graphs on the right hand side show the minimal potential energy of the two atoms depending on the atomic distance in $Z$-direction, $Z_D$ (we still assume symmetric displacements in $Z$, i.e., $Z_S=0$). As long as the equilibrium position on the $Z$-axis is a local potential minimum, the minimal energy concordantly exhibits an energetic barrier towards smaller distances of the atoms. The peak of this energetic barrier is located at one of the two saddle points of the potential that are located symmetrically with respect to the longitudinal axis. When the local potential minimum is lost, these saddle points collapse into a single saddle point at the equilibrium position on the $Z$-axis where the local minimum simultaneously vanishes. This can be seen in the series of plots on the left hand side. The energy plots on the right hand side show that the energetic barrier also simultaneously vanishes.

We find the minimal values of the potential energy at positions for which $X_S$ is two orders of magnitude smaller than $X_D$, but nonzero. This is due to the permanent electric dipole moments $\bm d_\gamma$ for vanishing electric field, which in turn are a signature of the finite size of Rydberg atoms. The origin of these moments $\bm d_\gamma$ is described in App.~\ref{c:electricdipolemoment_zerofield} and their properties in the $X$-$Z$-plane are depicted in the left subfigure of Fig.~\ref{fig:dipoles}. The illustration shows that they point in the same direction as the electric dipole moments induced by the electric field ($\bm d_F$, middle plot) for positive displacements of the atoms in $X$. They are anti-parallel for negative $X$ and they vanish on the $Z$-axis. Since $\bm d_\gamma$ is significantly smaller than $\bm d_F$ for all considered electric fields in this chapter, $\bm d_\gamma$ can be considered a correction to $\bm d_F$ and their sum $\bm d$ is parallel to $\bm d_F$ but larger or smaller in magnitude than $\bm d_F$ for positive or negative $X$, respectively. This introduces an asymmetry in $X_S$ and $X_D$ into the dipole-dipole interaction energy and is hence responsible for the nonzero values of $X_S$ for the position of minimal potential energy. This asymmetry is also responsible for the negative values of $X_D$ for the positions of minimal potential energy shown in Figs.~\ref{fig:implosionGmax}.

The stability of the equilibrium configuration is insensitive to changes in the electric field strength. The same is true for the transversal part of the center of mass wave function of the atoms. We also note that changing the principal quantum number $n$ has no considerable effect on the stability of the equilibrium position as long as the requirements involving $n$ can be met. The stability of the equilibrium configuration of the two atoms hence
depends essentially on the magnetic field parameters $B$, $G$ and $Q$ but not significantly on $F$ and $n$.

In order to find an analytical stability condition involving the magnetic field parameters we examine the curvature of the potential in $X_D$-direction at the equilibrium position. For small ratios $G/B$ (which is the case when the system is close to collapse), and assuming $F\gg nG$, we find the approximate expression for the curvature of the dipole-dipole interaction energy in $X_D$-direction 
\begin{equation}
\left(  \frac{\partial^2}{\partial X_D^2} V_{\text{dd}} \right) \bigg|_\text{$Z$-axis} 
\approx \left( \frac{F_x}{B} \right)^2 \frac{n^6}{Z_D^5} \ ,
\end{equation}
which strongly depends on the distance of the atoms. Inserting the equilibrium distance $Z_{D,\text{min}}$ from Eq.~(\ref{eq:approxZDmin}), where the dipole-dipole repulsion and the longitudinal confinement add to zero, we find
\begin{equation}                                \label{e:approxcurvatureVdd}
  \left( \frac{\partial^2}{\partial X_D^2} V_{\text{dd}} \right) \bigg|_\text{equilibrium} 
  \approx -6nQ \ .
\end{equation}
For a stable configuration this anti-confinement has to be weaker than the transversal confinement due to the magnetic field at the equilibrium position, yielding the condition 
\begin{equation}                                 \label{e:stabilitycondition}
  \frac{G^2}{B^2} > 14 \tilde Q \ .
\end{equation}
As soon as the right-hand side of this inequality becomes as large as the left-hand side the two saddle points at the potential barriers join on the $Z$-axis and the local minimum is lost. Inserting the geometry parameter for the millimeter trap in Ref.~\cite{Moore}, $\tilde Q=6\times 10^{-16}$, the stability condition takes the explicit form $G\,[\mathrm{Tm}^{-1}]  > 0.17\, B\,[\mathrm{Gauss}]$.

The previously derived condition (\ref{e:condmagnconf}) is less restrictive than (\ref{e:stabilitycondition}) for all presented examples. This means that the collapse is interaction-induced and it is not due to the loss of magnetic confinement. 

For the existence of a stable configuration of the two atoms it does, in fact, not suffice to have a local minimum on the $Z$-axis, i.e., to meet the requirement (\ref{e:stabilitycondition}). The potential around the local minimum must additionally be deep enough to accommodate at least the two-atom center of mass ground state. The height of the deciding potential barrier can be increased by tightening the transversal confinement. For the exemplary parameter set \B{1}, \G{1}, \F{10}, $Q=6\times 10^{-16}B$, the \cm~ground state energy is $E_0=13.9\,$kHz, and the barrier height is $2140\,$kHz $\mathrel{\widehat{=}}51.4\,\mu$K. This temperature can typically be reached in a magneto-optical trap without further cooling.

%%%%%%%%%%%%%%% 
\section{Excitation schemes\label{section:exschemes}}

The techniques that have been suggested and used to excite atoms into circular Rydberg states range from the microwave transfer method of Hulet and Kleppner \cite{Hulet1983}, the crossed-fields method proposed by Delande and Gay \cite{Delande1988}, and the RF field method proposed by Molander et al.~\cite{Molander1986}. The concern of this section, however, is not the excitation of single atoms into circular states but the excitation of two atoms in a Ioffe-Pritchard trap with an additional electric field directly into the equilibrium configuration that is stabilized by the dipolar interaction of the atoms. Solutions to this complication include (\emph{i}) externally forcing the excitation to happen at the desired positions only, or (\emph{ii}) changing the single-atom potential such  that its minima coincide with the two-atom potential minimum, or even (\emph{iii}) adiabatically transferring the system to the desired state by varying the detuning of the laser during the excitation process. A possible implementation of each of the options is described in the following. In principle all schemes are extendable to more than two atoms.

The only coordinates that have to be externally imposed are the $Z$-coordinates of the atoms since zero transversal displacement minimizes the energy for both the single-atom as well as the two-atom potential. The cleanest way to do this is to trap individual ground state atoms in two optical dipole traps at the desired $Z$-positions. The trapping volume of such optical tweezers \cite{Ashkin1986} can be made small enough (less than a $\mu$m in diameter) that only one atom can be captured in each trap \cite{Schlosser2001}. These two atoms can then be excited using one of the methods described in Ref.~\cite{Hulet1983}.

Another way of forcing the Rydberg atoms to be produced at the desired $Z$-positions is to excite them from a cold ground state atomic cloud by two laser beams perpendicular to the $Z$-axis that are focused next to each other to the desired equilibrium positions of the Rydberg atoms. This is possible if the equilibrium distance is considerably larger than the waist of the focused laser beams, which can be as small as one $\mu$m. Due to the strong Rydberg-Rydberg interaction, which yields an energy shift within the excitation volume that is larger than the linewidth of the laser, only one atom can be excited within one of the laser beams. As the excitation can be located at any of the atoms in that region, however, the ensemble of atoms is excited collectively into a superposition state called \emph{superatom} \cite{Vuletic2006}.

The second solution involves the modification of the single Rydberg atom potential. This can be done by adding an extra wire on the $X$-axis to the Z-trap on a chip or, correspondingly, by adding an extra coil between the coils of a macroscopic Ioffe-Pritchard trap that are responsible for the Ioffe field. Both setups yield a double well potential with a variable barrier height and a variable distance of the potential minima. For vanishing electric field two Rydberg atoms can be excited independently from each other, one in the bottom of each well, by tuning the laser just under the energy of the minimum. In order to keep heating as low as possible, the magnetic barrier can now be substituted by the dipolar repulsion between the atoms by decreasing the current through the extra coil or wire and simultaneously increasing the electric field strength.

The circularization of the Rydberg atoms with a modified adiabatic rapid passage method \cite{Nussenzveig1993}, for example, can be completed within $5\,\mu$s. The timescale of changing the magnetic field strongly depends on the configuration. If it is small enough, the described excitation scheme is scalable to produce more than two excitations, i.e., a Rydberg atom chain. This can be done by applying a magnetic field gradient in $Z$-direction which tilts the trap and moves the stable Rydberg atom pair in $Z$-direction. The magnetic barrier can be ramped up again as to confine the pair in one of the wells. At the minimum of the other well an additional circular Rydberg atom can be produced. The two atoms in the first well mutually tune themselves out of resonance of the exciting laser due to their interaction. By ramping the magnetic barrier up and down, and exciting a Rydberg atom in the empty well every time as described, a stable chain of atoms can be produced in the trap. The procedure is restricted by the timescales of excitation, magnetic field switching and by the lifetime of the circular Rydberg atoms which is about $2\,$ms for $n=30$ in the field-free case and scales with $n^5$.

With both schemes mentioned above, the atoms are excited into single-atom potential minima whose positions have to match the minimum of the two-atom potential that includes the interaction. Instead of artificially creating single-atom potential minima outside the origin, one can adiabatically transfer a cloud of ground state atoms from the state with no excitation via the state with one excited atom at the origin to the stable equilibrium state for two atoms relying on the structuring effect of the dipole-dipole repulsion. The electric field must thus be switched on at the beginning of the procedure.

The idea is based on the dynamical crystallization approach of Pohl et al.~\cite{Pohl2010}. The starting point is a cold gas of ground state atoms that can be modeled as consisting of two-level systems. The two levels considered in Ref.~\cite{Pohl2010} are the ground and the low angular momentum $nS_{1/2}$ Rydberg state. Here, of course, we need the excited level to be the circular Rydberg state, which can be achieved by utilizing the rf-optical excitation technique described by Cheng et al.~\cite{Cheng1994}. Within the dynamical crystallization scheme, the coupling laser is detuned against the two-photon resonance. For large negative detunings the many-body ground state in the rotating frame of reference coincides with the initial state where all atoms are in the ground state. Increasing the detunings to positive values effectively lowers the energy levels of many-body states with one and two and more excitations. They cross at critical detunings $\Delta_1^0$, $\Delta_2^1$, \dots , and states with 1, 2 and more Rydberg atoms are populated. The detuning is hence a control parameter that decreases the energy difference of adjacent number states $\ket{0}$, $\ket{1}$, $\ket{2}$,\dots\ with zero, one, two\dots\ excitations, respectively \cite{Tezak2010}. Since the laser couples the different number states, their energies undergo avoided crossings of separations $\delta_1^0$, $\delta_2^1$,\dots\ at the critical detunings. An adiabatic preparation of the states  $\ket{1}$, $\ket{2}$,\dots\ is possible as long as the time in which the detuning of the laser changes is large compared to $1/\delta_1^0$, $1/\delta_2^1$,\dots .

At the first crossing the initial state $\ket{0}$, with all atoms in the ground state, is directly coupled to the first excited state $\ket{1}$, with one Rydberg atom at the origin, which yields $\delta_1^0\sim \Omega$. From $\ket{1}$ to $\ket{2}$, however, there is no direct laser coupling since the energetically lowest state with two Rydberg atoms, $\ket{2}$, is the stable equilibrium configuration described in Sec.~\ref{c:1Dstableconfiguration} with two Rydberg atoms symmetrically displaced from the origin. To go from $\ket{1}$ to $\ket{2}$, two off-resonant intermediate steps are required. First the central atom is de-excited and subsequently the two Rydberg atoms at their equilibrium position are excited. A three-photon process is hence needed to come from $\ket{1}$ to $\ket{2}$. Assuming that $\ket{1}$ and $\ket{2}$ are resonant at time $t$, then the intermediate states are detuned by $\Delta\Omega_i$. For the parameters used in Section~\ref{c:1Dstableconfiguration} (e.g., in Fig.~\ref{f:secaroundlocmin2}; \B{30}, \G{10}, $Q=6\times 10^{-16}B$, $F=2\times 10^{-11}$\,a.u., $n=30$) all detunings $\Delta\Omega_i$ are of the order of $\sim 100$~kHz. If the Rabi frequency is much smaller than the intermediate state detunings $\Delta\Omega_i$, then the intermediate states that couple  $\ket{1}$ and $\ket{2}$ act as virtual levels for a resonant multi-photon transition. For larger Rabi frequencies, however, $\Omega (t) >\Delta\Omega_i$, power broadening exceeds the intermediate state detunings and the states are coupled by consecutive one photon transitions.

%%%%%%%%%%%%%%%%%%%%%%%%%%%%%%%%%%%%%%%%%%%%%%%%%%%%%%%%%%%%%%%%%%%%%%%%%%%%%%%%%%%%%%%%%%%%%% 
%%%%%%%%%%%%%%%%%%%%%%%%%%%%%%%%%%%%%%%%%%%%%%%%%%%%%%%%%%%%%%%%%%%%%%%%%%%%%%%%%%%%%%%%%%%%%%
\section{Brief summary\label{section:summary}}

In the present work we investigated the controlled trapping of two individual Rydberg atoms by means of a magnetic Ioffe-Pritchard trap that is superimposed by a constant electric field. The single-atom adiabatic potentials for such a field configuration have been derived and discussed. Including the interaction of the two Rydberg atoms, analytic expressions for the equilibrium positions of the two involved Rydberg atoms could be derived in the regime of a strong transversal confinement. As an interesting result, it turned out that the distance between the two atoms can be easily tuned without altering the involved trap frequencies by changing the applied electric field.
Loosening the restriction of a strong transversal confinement, on the other hand, leads to truly three-dimensional potential surfaces that, in principle, allow for the collapse of the system. The regime of stable trapping has been identified and the resulting adiabatic potentials were discussed.
Possible routes to experimentally realize the proposed system have been outlined.

\begin{acknowledgments}
B.H. would like to thank Igor Lesanovsky and Peter {Kr{\"u}\-ger} for fruitful discussions. 
This work was supported by the Deutsche Forschungsgemeinschaft (DFG) within the framework of the Excellence Initiative through the Heidelberg Graduate School of Fundamental Physics (GSC 129/1). P.S. acknowledges support by the DFG.
\end{acknowledgments}

\appendix
\section{Computing electric dipole moment expectation values perturbatively\label{appendix:PT}}

In order to be able to base the perturbative treatment in the uppermost surface on the circular Rydberg state, the perturbation operator has to be rotated into the local direction of the magnetic field.

\subsection{{Permanent electric dipole moments as finite size effect}\label{c:electricdipolemoment_zerofield}} 
In the analytically diagonalizable case of high Ioffe field the electronic state corresponding to the uppermost electronic adiabatic energy surface is the circular state with respect to the local field direction as quantization axis. For a finite gradient $G$ the purely electronic finite-size term $\mathcal H_\gamma$ admixes states to the circular state that have opposite parity. The total wave function thus looses its definite parity and the matrix elements of the odd dipole operator $e\bm r$ no longer vanish identically due to symmetry. This can result in a permanent electric dipole moment.

Because the unperturbed state vector is analytically given in the rotated system, i.e., with the local magnetic field axis as the quantization axis, the perturbation operator has to be rotated into this local frame, too. It is convenient to additionally replace the momentum operator in $\mathcal H_\gamma$ by angular momentum operators. This can be done exploiting the energetic degeneracy of an $n$-manifold in a field-free environment. The commutator $[xyz, H_A]$ vanishes within an $n$-manifold, it is therefore $\bra{\varphi} xyp_z \ket{\varphi} = \frac{1}{3} \bra{\varphi}(xL_x -yL_y)\ket{\varphi}$, and the perturbation operator in the transformed frame becomes
\begin{align}                                                  \label{e:gammatransformed}
  W &= U\mathcal H_\gamma U^\dagger  
  =U \lambda (xL_x -yLy +3xS_x -3yS_y) U^\dagger \nonumber \\
  &=  \lambda \{
  (\mathcal R^{-1}_{\alpha\beta} \bm r)_x \cdot (\mathcal R^{-1}_{\alpha\beta} \bm L)_x 
  -(\mathcal R^{-1}_{\alpha\beta} \bm r)_y \cdot (\mathcal R^{-1}_{\alpha\beta} \bm L)_y  \nonumber\\
  &\phantom{=} +3 (\mathcal R^{-1}_{\alpha\beta} \bm r)_x \cdot (\mathcal R^{-1}_{\alpha\beta} \bm S)_x 
  -3 (\mathcal R^{-1}_{\alpha\beta} \bm r)_y \cdot (\mathcal R^{-1}_{\alpha\beta} \bm S)_y  \} \nonumber\\
  &=: \lambda \ c_{ij}(\bm R,B,G,Q)\ r_i (L_j +3 S_j) \ ,
\end{align}
where $\lambda=G/3$~a.u.\ and $\mathcal R_{\alpha\beta}$ is the rotation associated with the transformation $U$, cf.\ Eq.~(\ref{eq:diagtrafoQ}). We write $\mathcal R^{-1}_{\alpha\beta}$ in Eq.~(\ref{e:gammatransformed}) instead of $\mathcal R_{\alpha\beta}$ to recall the fact that the components of a \emph{vector operator} transform in the rotation $\mathcal R$ like those of a vector in the rotation $\mathcal R^{-1}$ \cite{Messiah2}. Both the coordinate vector $\bm r$ and the angular momentum operators $\bm L$ and $\bm S$ are vector operators. 

We note that the perturbation operator, which is of purely electronic nature in the laboratory frame, now depends on the center of mass coordinates $\bm R$ through the coefficients $c_{ij}(\bm R,B,G,Q)$ in the rotated frame.

The \emph{first order energy correction} to the uppermost circular state $\ket{\psi_c}$ vanishes,
\begin{align}                                                \label{e:lambdaepsilon1}
  \lambda\epsilon_1
  &= \braopket{\psi_c}{\lambda c_{ij}r_i (L_j +3S_j)}{\psi_c} = 0 \ .
\end{align}
The \emph{first order correction in the wave function} reads
\begin{align}                                                                    \label{eq:lamda1def}
  \ket{\lambda \varphi_1^{(1)}}
  &= \sum_{p\ne 1} \frac{\braopket{\varphi_p}{\lambda c_{ij}\ r_i (L_j +3S_j)}{\varphi_1}}{E_1^0-E_p^0} \ket{\varphi_p} 
  \nonumber \\
  &=\mathrel{\mathop:} \sum_{p\ne 1}  \frac{W_{p1}}{E_{1p}} \ket{\varphi_p}
  =\mathrel{\mathop:} \sum_{p\ne 1}  f_p \ket{\varphi_p} \ .
\end{align}
Here we introduced the abbreviations $E_{pq}=E_p^0-E_q^0$ and $W_{pq}=\braopket{\varphi_p}{\lambda c_{ij}r_i(L_j +3S_j)}{\varphi_q}$, and we use the symbols $\ket{\varphi} =\ket{n,l,m,m_s}$ for the unperturbed hydrogenic electronic states in energetic order, starting with the circular state constituting the uppermost surface,
\begin{align}                                               \label{e:naminghydrogeniceigenfunctions}
\ket{\psi_c}= \ket{\varphi_1}=& \ket{n,n-1,n-1,\half} \ , \nonumber\\
\ket{\varphi_2}=& \ket{n,n-1,n-2,\half} \ , \nonumber\\
\ket{\varphi_3}=& \ket{n,n-2,n-2,\half} \ , \nonumber\\
\ket{\varphi_4}=& \ket{n,n-1,n-3,\half} \ , \nonumber\\
\ket{\varphi_5}=& \ket{n,n-2,n-3,\half} \ , \nonumber\\
\ket{\varphi_6}=& \ket{n,n-3,n-3,\half} \ , \nonumber\\
\ket{\varphi_7}=& \ket{n,n-1,n-1,-\half} \ , \nonumber\\
\vdots \phantom{===}& \nonumber\\
\ket{\varphi_{13}}=& \ket{n,n-2,n-2,-\half} \ .
\end{align}  
The states $\{\ket{\varphi_2}, \ket{\varphi_3}\}$, the states $\{\ket{\varphi_4}, \dots , \ket{\varphi_7}\}$  and the states $\{\ket{\varphi_8}, \dots , \ket{\varphi_{13}}\}$  are energetically degenerate in the limit  $B/G\rightarrow \infty$. The quantum numbers are given with respect to the local quantization axis which is the direction of the magnetic field. 

In order to compute the matrix elements $W_{p1}=W_{p,\psi_c}$ defined in Eq.~(\ref{eq:lamda1def}) we rewrite the angular momentum operators with ladder operators, 
\begin{align}
  L_x \ket{\psi_c}&=
  \frac{1}{2}(L_++L_-) \ket{\psi_c}=
  \frac{1}{2}L_- \ket{\psi_c},\\
  L_y \ket{\psi_c}&=  \label{eq:l_y_is_iL_xnew}
  \frac{1}{2i}(L_+-L_-) \ket{\psi_c}=
  i L_x \ket{\psi_c}, 
\end{align}
where $L_-\ket{l,m} = \sqrt{l(l+1)-m(m-1)} \ \ket{l,m-1}$, ($\hbar=1$~atomic unit). The only non-vanishing matrix elements in (\ref{eq:lamda1def}) are
\begin{align}                                        \label{eq:ME_lm_cijriLj_circnew}
  \braopket{^{\varphi_3}_{\varphi_5}}{c_{ix}r_i L_{x}}{\psi_c}& 
  =  \frac{1}{2}\sqrt{2n-3} \braopket{^{\varphi_3}_{\varphi_5}}{c_{ix}r_i}{\varphi_2}\ ,\\
  \braopket{^{\varphi_3}_{\varphi_5}}{{c_{iy}r_i L_{y}}}{\psi_c}& 
  = i \frac{1}{2}\sqrt{2n-3} \braopket{^{\varphi_3}_{\varphi_5}}{c_{iy}r_i}{\varphi_2}\ ,\\
  \braopket{\varphi_3}{c_{iz} r_i L_z}{\psi_c\psi_}& =  (n-1) \braopket{\varphi_3}{c_{iz}r_i}{\psi_c} 
\end{align}
due to the dipole selection rules $\Delta l=\pm 1$, $\Delta m_l=0,\pm 1$, and $\Delta m_s=0$. We proceed similarly with the spin operators, 
\begin{align}
  S_x\ket{m_s=\pm\frac{1}{2}} &= \frac{1}{2}\ \ket{m_s=\mp\frac{1}{2}} \quad\text{and} \\
  S_y\ket{m_s=\pm\frac{1}{2}} &= \pm\frac{i}{2}\ \ket{m_s=\mp\frac{1}{2}} \ .
\end{align}
The only non-vanishing matrix elements involving the spin operators are
\begin{align}                                        \label{eq:ME_lm_cijriSj_circnew}
  \braopket{\varphi_{13}}{c_{ix}r_i S_{x}}{\psi_c}& 
  =  \frac{1}{2} \braopket{\varphi_{13}}{c_{ix}r_i}{\varphi_7}\ ,\\
  \braopket{\varphi_{13}}{c_{iy}r_i S_{y}}{\psi_c}& 
  = \frac{i}{2} \braopket{\varphi_{13}}{c_{iy}r_i}{\varphi_7}\ ,\\
  \braopket{\varphi_3}{c_{iz} r_i S_z}{\psi_c}& =  \frac{1}{2} \braopket{\varphi_3}{c_{iz}r_i}{\psi_c} \ .
\end{align}
Considering the following relations between the dipole matrix elements, 
\begin{align}                                                                           \label{eq:xymenew}
  \braopket{l^\prime,m^\prime}{y}{l,m}& = \pm i\ \braopket{l^\prime,m^\prime}{x}{l,m}&  \text{for}\  m^\prime= & m\mp 1 \nonumber\\
  \braopket{l^\prime,m^\prime}{y}{l,m}& = 0 =\ \braopket{l^\prime,m^\prime}{x}{l,m}&    \text{for}\  m^\prime= & m \nonumber\\
  \braopket{l^\prime,m^\prime}{z}{l,m}& \sim \delta_{m,m^\prime} \ , &&
\end{align}
we find the first order correction to the wave function, 
\begin{align}                                                                    \label{eq:lamda1new}
  &\ket{\lambda \varphi_1^{(1)}} 
  \!=\!\bigg\{\!
  (c_{xz}+ic_{yz})   \frac{\ket{\varphi_3}}{E_{13}} 
  \!\left(\! (n-\frac{1}{2})x_{31} +\frac{1}{2}\sqrt{2n-3}z_{32} \!\right)\nonumber\\ 
  &\phantom{\ket{\lambda \varphi_1^{(1)}} 
  \!=\!\bigg\{\!\! }
  +\frac{\ket{\varphi_5}}{E_{15}}  (c_{xx}-c_{yy}+2ic_{xy})  \frac{1}{2}\sqrt{2n-3} x_{52}\nonumber\\ 
  &\phantom{\ket{\lambda \varphi_1^{(1)}} 
  \!=\!\bigg\{\!\! }
  +\frac{\ket{\varphi_{13}}}{E_{13,5}}  (c_{xx}-c_{yy}+2ic_{xy})  \frac{3}{2} x_{13,7}
  \bigg\} \lambda 
\ .
\end{align}
Here we introduced the notation $x_{ij}= \braopket{\varphi_i}{x}{\varphi_j}$, $y_{ij}= \braopket{\varphi_i}{y}{\varphi_j}$ and $z_{ij}= \braopket{\varphi_i}{z}{\varphi_j}$. The following explicit expressions for the matrix elements can be deduced from the formulas for the radial and angular integrals involving hydrogenic wave functions in~\cite{Zimmermann1979},
\begin{align}
  z_{32}& =-\frac{3}{2}n \ , \\
  x_{13}& =\   \frac{3}{2\sqrt{2}} n \sqrt{n-1} \ , \\  
  x_{52}& =\   \frac{3}{2\sqrt{2}} n \sqrt{n-2} \ ,  \\
  x_{13,7}& =\ \frac{3}{2\sqrt{2}} n \sqrt{\frac{(n-1/2)(n-3/2)}{n-2}} \ .
\end{align}
Note that the correction $\ket{\lambda \varphi_1^{(1)}}$ to the circular wave function $\ket{\varphi_1}=\ket{\psi_c}$ has definite parity since $\ket{\varphi_3}$, $\ket{\varphi_5}$ and $\ket{\varphi_{13}}$ have the same $l$ quantum number. It is opposite to the parity of $\ket{\psi_c}$, however. The involved coefficients $c_{ij}(\bm R,B,G,Q)$, defined in Eq.~(\ref{e:gammatransformed}), come from the inverse rotation of $\bm r$, $\bm L$ and $\bm S$ with $\mathcal R_{\alpha\beta}$. Expressed via the rotation angles $\alpha$ and $\beta$, the coefficients read explicitly
\begin{align}    \label{eq:cijfirstversionnew}
  c_{xx} =& \cos^2\beta \ , \nonumber \\
  c_{yx} =& \sin\alpha\sin\beta\cos\beta \ , \nonumber \\
  c_{zx} =& -\cos\alpha\sin\beta\cos\beta \ , \nonumber \\
  c_{xy}=& \sin\alpha\sin\beta\cos\beta \ , \nonumber \\
  c_{yy}=& (\sin^2\alpha\sin^2\beta-\cos^2\alpha)  \ , \nonumber \\
  c_{zy}=& - \sin\alpha\cos\alpha (1+\sin^2\beta) \ , \nonumber \\
  c_{iz} =& -\cos\alpha \sin\beta\cos\beta  \ , \nonumber \\
  c_{yz} =& -\sin\alpha\cos\alpha(1+\sin^2\beta)  \ , \nonumber \\
  c_{zz} =& (\cos^2\alpha\sin^2\beta-\sin^2\alpha) \ , 
\end{align}
where $c_{ij}=c_{ji}$. This is a general expression for the perturbation operator (\ref{e:gammatransformed}). The particular magnetic field configuration only enters via the explicit expressions for the angles $\alpha$ and $\beta$. On the $Z$-axis ($\alpha=\beta=0,\,U=\bm 1$) all the coefficients but $c_{xx}=1$ and $c_{yy}=-1$ vanish. The correction to the circular state in first order reduces to 
\begin{align}                              \label{eq:lambda1originnew}
  \ket{\lambda \varphi_1^{(1)}} (O)
  \approx & \lambda \frac{3}{2}n\sqrt{n} 
  \left( 
    \sqrt{n} \frac{\ket{\varphi_5}}{E_{51}}
    +\frac{3}{\sqrt{2}}\frac{\ket{\varphi_{13}}}{E_{13,5}}   
  \right)  
\end{align}
for large $n$. The electric dipole moment expectation value therefore vanishes at the origin due to the dipole selection rules as will be described in the following.

%% Electric Dipole Moment Expectation Value %%%%%%%%%%%%%%%%%%%%%%%%%%%%%%%%%%%%%%%%%%%%
The \emph{electric dipole moment} of the electronic state to second order in perturbation theory is found by
computing the expectation value of the dipole operator rotated into the local direction of the magnetic field, $U^\dagger\bm r U$, for the perturbed state in the rotated frame, $\ket{\varphi_1 +\lambda\varphi_1^{(1)}}$,
\begin{align}                                  \label{eq:Dfirstordercalcnew}
  \bm d_\gamma = & \braopket{\varphi_1 +\lambda\varphi_1^{(1)}}{U^\dagger\bm r U}{\varphi_1 +\lambda\varphi_1^{(1)}} \nonumber\\
  =  &\braopket{\varphi_1}{U^\dagger\bm r U}{\varphi_1}
  +\braopket{\lambda\varphi_1^{(1)}}{U^\dagger\bm r U}{\lambda\varphi_1^{(1)}} \nonumber\\
  & +2\ \text{Re} (\braopket{\varphi_1}{U^\dagger\bm r U}{\lambda\varphi_1^{(1)}}) \ .
\end{align}
The second line in Eq.~(\ref{eq:Dfirstordercalcnew}) vanishes due to the definite parity of $\ket{\varphi_1}$ and $\ket{\lambda\varphi_1^{(1)}}$ and the matrix element in the third term simplifies due to the dipole selection rules (since $\Delta m=2$ for $\braopket{\varphi_1}{U \bm r U^\dagger}{\varphi_5}$ and $\Delta m_s=1$ for $\braopket{\varphi_1}{U \bm r U^\dagger}{\varphi_{13}}$),
\begin{align}
  \braopket{\varphi_1}{U^\dagger\bm r U}{\lambda\varphi_1^{(1)}}
  &=f_3 U^\dagger \braopket{\varphi_1}{\bm r}{\varphi_3} U
  =f_3 \mathcal R_{\alpha\beta}^{-1} \left(\!\!\!\begin{array}{c}x_{13}\\-ix_{13}\\0\end{array}\!\!\!\right), \nonumber
\end{align}
and it is
\begin{equation}                                         \label{e:Dgammarotation}
  \bm d_\gamma 
  =2\ \text{Re} \!\left(\!
    f_3 \mathcal R_{\alpha\beta}^{-1} \left(\!\!\begin{array}{c}x_{13}\\-ix_{13}\\0\end{array}\!\!\right) \!\!\right) \!\!
  =  \lambda\chi 
  \mathcal R_{\alpha\beta}^{-1} \left(\!\!\begin{array}{c}c_{xz}\\c_{yz}\\0\end{array}\!\!\right),
\end{equation}
where $\chi:=9n^2(2n^2 -3n -\sqrt{4n^2 -10n +6} +1) /(8\Delta E)$ and $\lambda=G/3$. We can find the explicit spatial dependence of the electric dipole moment by expressing the angles $\alpha$ and $\beta$ with the magnetic field components $B_i$, to find the expression~(\ref{e:Dgammaspacialdependencesim}). For a Ioffe-Pritchard magnetic field configuration, $\bm d_\gamma $ vanishes on the $Z$-axis because the magnetic field components $B_x$ and $B_y$ are zero there. It is convenient for the symmetry analysis to also write down the explicit form of $\bm d_\gamma$ for $Q=0$,
\begin{equation}
  \bm d_{\gamma}(Q=0)
  =\lambda\chi\frac{G^3}{|\bm B|^{3}} 
  \left(
    \begin{array}{c}
      X \left( 2Y^2 +{B}^2/{G}^2\right) \\
      Y \left( 2X^2 +{B}^2/{G}^2\right) \\
      \left(Y^2-X^2\right)   \left({B}/{G}\right) 
    \end{array}
  \right) \ .
\end{equation}

%% perpendicular %%%%%%%%%%%%%%%%%%%%%%%%%%%%%%%%%%%%%%%%%%%%%%%%%%%%%%%
As can be deduced from Eq.~(\ref{e:Dgammarotation}), the electric dipole moment expectation value is perpendicular to the local direction of the magnetic field,
\begin{equation}                                                          \label{e:dgammaperpB}
  \bm d_\gamma \cdot \bm B \sim 
  \mathcal R_{\alpha\beta}^{-1} \left(\begin{array}{c} c_{xz} \\ c_{yz} \\0\end{array}\right) \cdot \bm B =0 \ .
\end{equation}

%% Section %%%%%%%%%%%%%%%%%%%%%%%%%%%%%%%%%%%%%%%%%%%%%%%%%%%%%%%
\subsection{{Non-parallel moments in an electric field}\label{c:electricdipolemoment_finitefield}} 
The Hamiltonian for the additional external electric field is $H_F=q\phi =(xF_x+yF_y+zF_z)$ since $\bm F= -\text{grad}\ \phi$ and $q=-e$ ($=-1$ in atomic units). According to the considerations in the preceding chapter the perturbation operator therefore reads 
\begin{equation}  \label{eq:HFwithoutscaling}
  W_F 
  = U(\bm r \cdot\bm F)U^\dagger 
  = U\bm r U^\dagger \cdot\bm F
  =: c_{F,ij}F_i r_j \ ,
\end{equation}
where the small parameter in the operator $W_F$ is the modulus of the electric field, $\lambda_F=|\bm F|$. Considering the Zeeman term dependence, $\bm \mu\bm B\sim |\bm B|$, the perturbation parameter is the ratio of the field strengths, $\lambda_F=|\bm F|/|\bm B|$.

The \emph{first order energy correction} due to $W_F$ vanishes due to the dipole selection rules since $\Delta l =0$, 
\begin{equation}
  \lambda_F \epsilon^{(1,F)}
  =\braopket{\psi_c}{c_{F,ij}F_i r_j}{\psi_c}=0 \ .
\end{equation}

The \emph{first order correction to the circular state} is
\begin{align}                                                                    \label{eq:lambda1F}
  \ket{\lambda_F \varphi_1^{(1,F)}}
  & = \sum_{p\ne 1} \frac{\braopket{\varphi_p}{c_{F,ij}F_i r_j}{\varphi_1}}{E_1^0-E_p^0} \ket{\varphi_p} \nonumber\\
  & =\mathrel{\mathop:} \sum_{p\ne 1}  \frac{W_{F,p1}}{E_{1p}} \ket{\varphi_p}
  = \frac{W_{F,31}}{E_{13}} \ket{\varphi_3} \ ,
\end{align}
where we introduced $W_{F,pq}=\braopket{\varphi_p}{c_{F,ij}F_i r_j}{\varphi_q}$ and $E_{pq}=E_p^0-E_q^0$. For an electric field in arbitrary direction the numerator in~(\ref{eq:lambda1F}) reads 
\begin{align}     
  W_{F,31}  = e x_{31}  [ &
     F_x(c_{F,xx} +i c_{F,xy}) \nonumber\\
    &+F_y i c_{F,yy}
    +F_z(c_{F,zx} +i c_{F,zy})   ]
\end{align}
and if we restrict our consideration to an electric field pointing along the $X$-axis we find
\begin{equation}                                                                    \label{eq:lambda1Fx}
  \ket{\lambda_F \varphi_1^{(1,F_x)}}
  =  ex_{13} F_x(c_{F,xx} +i c_{F,xy})  \frac{\ket{\varphi_3}}{E_{13}} \ ,
\end{equation}
where $c_{F,xx}= \cos\beta$ and $c_{F,xy}= \sin\alpha\sin\beta$.

The \emph{second order energy correction} due to the external electric field reads
\begin{align}
 \lambda_F^2 \epsilon^{(2,F_x)}
 = & F_x\braopket{\varphi_1+\lambda_F\varphi_1^{(1,F_x)}}{c_{F,xj}r_j}{\varphi_1+\lambda_F\varphi_1^{(1,F_x)}} \nonumber\\
 = & 2 \text{Re} \left( F_x\braopket{\varphi_1}{c_{F,xj}r_j}{\lambda_F\varphi_1^{(1,F_x)}}   \right)\nonumber\\
 = & 2\frac{e F_x^2}{E_{13}} x_{13}^2 (c_{F,xx}^2 +c_{F,xy}^2)  \nonumber\\
 = & \frac{9}{4}\frac{F_x^2}{E_{13}} n^2(n-1) (\cos{\beta}^2 +\sin{\alpha}^2 \sin{\beta}^2) \ .
\end{align}
For vanishing $Q$ and with the approximate expression for the energetic separation between the coupling surfaces, $E_{13}:=\Delta E \approx |\bm B|/2$, this reads
\begin{equation}                                                       
 \lambda_F^2 \epsilon^{(2,F_x)}
 \approx  \frac{9}{4} F_x^2 n^2(n-1) \frac{B^2+G^2Y^2 \phantom{+G^2X^2}}{B^2+G^2Y^2+G^2X^2} \ .
\end{equation}
The perturbative contribution to the uppermost surface due to an external electric field $(F_x,0,0)$ is thus positive and it is maximal on the $z$ axis.

Both the unperturbed wave function $\ket{\varphi_1}$ as well as the perturbation $\ket{\lambda_F\varphi_1^{(1,F_x)}}$ have definite parity. The \emph{electric dipole moment expectation value} in the uppermost electronic energy surface is therefore, analogously to (\ref{eq:Dfirstordercalcnew}),
\begin{align}                                         \label{eq:Dfirstordercalc2F}
 &\bm d_F 
 =  2 \text{Re} \left(\braopket{\varphi_1}{U^\dagger\bm rU}{\lambda_F\varphi_1^{(1,F_x)}}   \right) \nonumber\\
  = & \mathcal R_{\alpha\beta}^{-1}
  (2x_{31} \frac{F_x}{E_{13}} \text{Re}(
  c_{F,x}  \left(\begin{array}{ccc} x_{13}\\-\imath x_{13}\\0  \end{array} \right)
  +c_{F,y} \left(\begin{array}{ccc} \imath x_{13}\\ x_{13}\\0  \end{array} \right))) \nonumber\\
  = &  \frac{2F_x}{E_{13}} x_{13}^2 
  \mathcal R_{\alpha\beta}^{-1} \left(\begin{array}{ccc}  c_{F,x}\\ c_{F,y}\\ 0 \end{array} \right) \nonumber\\
  = & \frac{2eF}{E_{13}} x_{13}^2 
  (
    \cos\beta   \!\left(\!\begin{array}{c}\cos\beta\\0\\\sin\beta\end{array}\!\right) \!  
   +\sin\alpha\sin\beta \!\left(\!\begin{array}{c}\sin\alpha\sin\beta\\\cos\alpha\\-\sin\alpha\cos\beta\end{array}\!\right) \!
  )  \nonumber\\
  = & \frac{9}{4}\frac{F_x}{E_{13}} n^2(n-1)
 \frac{1}{\bm B^2}\left(\begin{array}{ccc}  B_y^2+B_z^2\\ -B_xB_y\\ -B_xB_z \end{array} \right) 
\ ,
\end{align}
where we use $c_{F,x}=\cos{\beta}$, $c_{F,y}=\sin{\alpha}\sin{\beta}$, $x_{13}=\frac{3}{2\sqrt{2}}n\sqrt{n}$, 
and the relation (\ref{eq:xymenew}). In contrast to $\bm d_\gamma$, $\bm d_F$ does not vanish on the $Z$-axis 
but points in the direction of the electric field with $d_{F,x} =4x_{13}^2 {F_x}/{|\bm B|}$. This is not true away from the $Z$-axis. The dipole moment does not point in the electric field direction there but stays rather perpendicular to the local direction of the magnetic field [alike $\bm d_\gamma$, Eq.~(\ref{e:dgammaperpB})]. This can be deduced from the directional dependence in (\ref{eq:Dfirstordercalc2F}) for arbitrary magnetic field configurations,
\begin{equation}                                                          \label{e:dFperpB}
 \bm d_F \cdot \bm B  
 \sim \left(\begin{array}{ccc}  B_y^2+B_z^2\\ -B_xB_y\\ -B_xB_z \end{array} \right) \cdot \bm B
= 0 \ .
\end{equation}

\subsection{Addition of perturbatively calculated dipole moments}

For a non-zero external electric field the perturbation operator is $W_{\text{tot}} =U \mathcal H_\gamma U^\dagger + U \mathcal H_F U^\dagger$ and the total first order correction to the wave function hence reads 
$  \ket{\lambda 1} = \ket{\lambda_\gamma 1}_\gamma +\ket{\lambda_F 1}_F$. The calculation of the expectation value of the observable $\hat{O}$ in the perturbed state $\ket{\psi_c+\lambda 1}$ yields accordingly
\begin{align} 
  \braopket{\psi_c+\lambda 1}{&\hat{O}}{\psi_c+\lambda 1} \nonumber\\
  &=\hat{O}_\gamma+ \hat{O}_F  
  + 2 \text{Re}(\braopket{\lambda_\gamma 1_\gamma}{\hat{O}}{\lambda_F 1_F}) \ .
\end{align}
The perturbations $\ket{\lambda_\gamma 1}_\gamma$ and $\ket{\lambda_F 1}_F$, Eqs.~(\ref{eq:lamda1new}) and (\ref{eq:lambda1F}), involve the states $\ket{\varphi_{3}}$, $\ket{\varphi_{5}}$, $\ket{\varphi_{13}}$ which do not differ in their angular momentum quantum number $l$. They have the same definite parity since the parity of the spherical harmonics does not depend on the quantum number $m$. The mixed matrix element $\braopket{\lambda_\gamma 1_\gamma}{\bm d}{\lambda_F 1_F}$ therefore vanishes (the dipole operator is an odd operator) and 
\begin{equation}                                                                 \label{eq:d_gamma_plus_D_F}
 \bm d  = \braopket{\psi_c+\lambda 1}{\bm r}{\psi_c+\lambda 1} 
 = \bm d_\gamma+ \bm d_F \ .
\end{equation}
In other words, within first order perturbation theory, 
adding the different dipole moment expectation values $\bm d_\gamma$ and $\bm d_F$ 
is equivalent to calculating the expectation value for the combined perturbed state vector.

\bibliography{paper}

\begin{thebibliography}{46}%
\makeatletter
\providecommand \@ifxundefined [1]{%
 \@ifx{#1\undefined}
}%
\providecommand \@ifnum [1]{%
 \ifnum #1\expandafter \@firstoftwo
 \else \expandafter \@secondoftwo
 \fi
}%
\providecommand \@ifx [1]{%
 \ifx #1\expandafter \@firstoftwo
 \else \expandafter \@secondoftwo
 \fi
}%
\providecommand \natexlab [1]{#1}%
\providecommand \enquote  [1]{``#1''}%
\providecommand \bibnamefont  [1]{#1}%
\providecommand \bibfnamefont [1]{#1}%
\providecommand \citenamefont [1]{#1}%
\providecommand \href@noop [0]{\@secondoftwo}%
\providecommand \href [0]{\begingroup \@sanitize@url \@href}%
\providecommand \@href[1]{\@@startlink{#1}\@@href}%
\providecommand \@@href[1]{\endgroup#1\@@endlink}%
\providecommand \@sanitize@url [0]{\catcode `\\12\catcode `\$12\catcode
  `\&12\catcode `\#12\catcode `\^12\catcode `\_12\catcode `\%12\relax}%
\providecommand \@@startlink[1]{}%
\providecommand \@@endlink[0]{}%
\providecommand \url  [0]{\begingroup\@sanitize@url \@url }%
\providecommand \@url [1]{\endgroup\@href {#1}{\urlprefix }}%
\providecommand \urlprefix  [0]{URL }%
\providecommand \Eprint [0]{\href }%
\providecommand \doibase [0]{http://dx.doi.org/}%
\providecommand \selectlanguage [0]{\@gobble}%
\providecommand \bibinfo  [0]{\@secondoftwo}%
\providecommand \bibfield  [0]{\@secondoftwo}%
\providecommand \translation [1]{[#1]}%
\providecommand \BibitemOpen [0]{}%
\providecommand \bibitemStop [0]{}%
\providecommand \bibitemNoStop [0]{.\EOS\space}%
\providecommand \EOS [0]{\spacefactor3000\relax}%
\providecommand \BibitemShut  [1]{\csname bibitem#1\endcsname}%
\let\auto@bib@innerbib\@empty
%</preamble>
\bibitem [{\citenamefont {Gallagher}(1994)}]{gallagher94}%
  \BibitemOpen
  \bibfield  {author} {\bibinfo {author} {\bibfnamefont {T.~F.}\ \bibnamefont
  {Gallagher}},\ }\href@noop {} {\emph {\bibinfo {title} {Rydberg Atoms}}}\
  (\bibinfo  {publisher} {Cambridge University Press},\ \bibinfo {address}
  {Cambridge, U.K.},\ \bibinfo {year} {1994})\BibitemShut {NoStop}%
\bibitem [{\citenamefont {Singer}\ \emph {et~al.}(2004)\citenamefont {Singer},
  \citenamefont {Reetz-Lamour}, \citenamefont {Amthor}, \citenamefont
  {Marcassa},\ and\ \citenamefont {Weidem\"uller}}]{singer}%
  \BibitemOpen
  \bibfield  {author} {\bibinfo {author} {\bibfnamefont {K.}~\bibnamefont
  {Singer}}, \bibinfo {author} {\bibfnamefont {M.}~\bibnamefont
  {Reetz-Lamour}}, \bibinfo {author} {\bibfnamefont {T.}~\bibnamefont
  {Amthor}}, \bibinfo {author} {\bibfnamefont {L.~G.}\ \bibnamefont
  {Marcassa}}, \ and\ \bibinfo {author} {\bibfnamefont {M.}~\bibnamefont
  {Weidem\"uller}},\ }\href@noop {} {\bibfield  {journal} {\bibinfo  {journal}
  {Phys. Rev. Lett.}\ }\textbf {\bibinfo {volume} {93}},\ \bibinfo {pages}
  {163001} (\bibinfo {year} {2004})}\BibitemShut {NoStop}%
\bibitem [{\citenamefont {Tong}\ \emph {et~al.}(2004)\citenamefont {Tong},
  \citenamefont {Farooqi}, \citenamefont {Stanojevic}, \citenamefont
  {Krishnan}, \citenamefont {Zhang}, \citenamefont {C\^ot\'e}, \citenamefont
  {Eyler},\ and\ \citenamefont {Gould}}]{tong}%
  \BibitemOpen
  \bibfield  {author} {\bibinfo {author} {\bibfnamefont {D.}~\bibnamefont
  {Tong}}, \bibinfo {author} {\bibfnamefont {S.~M.}\ \bibnamefont {Farooqi}},
  \bibinfo {author} {\bibfnamefont {J.}~\bibnamefont {Stanojevic}}, \bibinfo
  {author} {\bibfnamefont {S.}~\bibnamefont {Krishnan}}, \bibinfo {author}
  {\bibfnamefont {Y.~P.}\ \bibnamefont {Zhang}}, \bibinfo {author}
  {\bibfnamefont {R.}~\bibnamefont {C\^ot\'e}}, \bibinfo {author}
  {\bibfnamefont {E.~E.}\ \bibnamefont {Eyler}}, \ and\ \bibinfo {author}
  {\bibfnamefont {P.~L.}\ \bibnamefont {Gould}},\ }\href@noop {} {\bibfield
  {journal} {\bibinfo  {journal} {Phys. Rev. Lett.}\ }\textbf {\bibinfo
  {volume} {93}},\ \bibinfo {pages} {063001} (\bibinfo {year}
  {2004})}\BibitemShut {NoStop}%
\bibitem [{\citenamefont {{Cubel Liebisch}}\ \emph {et~al.}(2005)\citenamefont
  {{Cubel Liebisch}}, \citenamefont {Reinhard}, \citenamefont {Berman},\ and\
  \citenamefont {Raithel}}]{Liebisch2005}%
  \BibitemOpen
  \bibfield  {author} {\bibinfo {author} {\bibfnamefont {T.}~\bibnamefont
  {{Cubel Liebisch}}}, \bibinfo {author} {\bibfnamefont {A.}~\bibnamefont
  {Reinhard}}, \bibinfo {author} {\bibfnamefont {P.~R.}\ \bibnamefont
  {Berman}}, \ and\ \bibinfo {author} {\bibfnamefont {G.}~\bibnamefont
  {Raithel}},\ }\href@noop {} {\bibfield  {journal} {\bibinfo  {journal} {Phys.
  Rev. Lett.}\ }\textbf {\bibinfo {volume} {95}},\ \bibinfo {pages} {253002}
  (\bibinfo {year} {2005})}\BibitemShut {NoStop}%
\bibitem [{\citenamefont {Vogt}\ \emph {et~al.}(2007)\citenamefont {Vogt},
  \citenamefont {Viteau}, \citenamefont {Chotia}, \citenamefont {Zhao},
  \citenamefont {Comparat},\ and\ \citenamefont {Pillet}}]{Vogt2007}%
  \BibitemOpen
  \bibfield  {author} {\bibinfo {author} {\bibfnamefont {T.}~\bibnamefont
  {Vogt}}, \bibinfo {author} {\bibfnamefont {M.}~\bibnamefont {Viteau}},
  \bibinfo {author} {\bibfnamefont {A.}~\bibnamefont {Chotia}}, \bibinfo
  {author} {\bibfnamefont {J.}~\bibnamefont {Zhao}}, \bibinfo {author}
  {\bibfnamefont {D.}~\bibnamefont {Comparat}}, \ and\ \bibinfo {author}
  {\bibfnamefont {P.}~\bibnamefont {Pillet}},\ }\href@noop {} {\bibfield
  {journal} {\bibinfo  {journal} {Phys. Rev. Lett.}\ }\textbf {\bibinfo
  {volume} {99}},\ \bibinfo {pages} {073002} (\bibinfo {year}
  {2007})}\BibitemShut {NoStop}%
\bibitem [{\citenamefont {van Ditzhuijzen}\ \emph {et~al.}(2008)\citenamefont
  {van Ditzhuijzen}, \citenamefont {Koenderink}, \citenamefont {Hern\'{a}ndez},
  \citenamefont {Robicheaux}, \citenamefont {Noordam},\ and\ \citenamefont
  {{van Linden van den Heuvell}}}]{Ditzhuijzen2008}%
  \BibitemOpen
  \bibfield  {author} {\bibinfo {author} {\bibfnamefont {C.~S.~E.}\
  \bibnamefont {van Ditzhuijzen}}, \bibinfo {author} {\bibfnamefont {A.~F.}\
  \bibnamefont {Koenderink}}, \bibinfo {author} {\bibfnamefont {J.~V.}\
  \bibnamefont {Hern\'{a}ndez}}, \bibinfo {author} {\bibfnamefont
  {F.}~\bibnamefont {Robicheaux}}, \bibinfo {author} {\bibfnamefont {L.~D.}\
  \bibnamefont {Noordam}}, \ and\ \bibinfo {author} {\bibfnamefont {H.~B.}\
  \bibnamefont {{van Linden van den Heuvell}}},\ }\href@noop {} {\bibfield
  {journal} {\bibinfo  {journal} {Phys. Rev. Lett.}\ }\textbf {\bibinfo
  {volume} {100}},\ \bibinfo {pages} {243201} (\bibinfo {year}
  {2008})}\BibitemShut {NoStop}%
\bibitem [{\citenamefont {Heidemann}\ \emph {et~al.}(2007)\citenamefont
  {Heidemann}, \citenamefont {Raitzsch}, \citenamefont {Bendkowsky},
  \citenamefont {Butscher}, \citenamefont {L\"{o}w}, \citenamefont {Santos},\
  and\ \citenamefont {Pfau}}]{Heidemann2007}%
  \BibitemOpen
  \bibfield  {author} {\bibinfo {author} {\bibfnamefont {R.}~\bibnamefont
  {Heidemann}}, \bibinfo {author} {\bibfnamefont {U.}~\bibnamefont {Raitzsch}},
  \bibinfo {author} {\bibfnamefont {V.}~\bibnamefont {Bendkowsky}}, \bibinfo
  {author} {\bibfnamefont {B.}~\bibnamefont {Butscher}}, \bibinfo {author}
  {\bibfnamefont {R.}~\bibnamefont {L\"{o}w}}, \bibinfo {author} {\bibfnamefont
  {L.}~\bibnamefont {Santos}}, \ and\ \bibinfo {author} {\bibfnamefont
  {T.}~\bibnamefont {Pfau}},\ }\href@noop {} {\bibfield  {journal} {\bibinfo
  {journal} {Phys. Rev. Lett.}\ }\textbf {\bibinfo {volume} {99}},\ \bibinfo
  {pages} {163601} (\bibinfo {year} {2007})}\BibitemShut {NoStop}%
\bibitem [{\citenamefont {Reetz-Lamour}\ \emph {et~al.}(2008)\citenamefont
  {Reetz-Lamour}, \citenamefont {Amthor}, \citenamefont {Deiglmayr},\ and\
  \citenamefont {Weidem\"{u}ller}}]{Reetz-Lamour2008}%
  \BibitemOpen
  \bibfield  {author} {\bibinfo {author} {\bibfnamefont {M.}~\bibnamefont
  {Reetz-Lamour}}, \bibinfo {author} {\bibfnamefont {T.}~\bibnamefont
  {Amthor}}, \bibinfo {author} {\bibfnamefont {J.}~\bibnamefont {Deiglmayr}}, \
  and\ \bibinfo {author} {\bibfnamefont {M.}~\bibnamefont {Weidem\"{u}ller}},\
  }\href@noop {} {\bibfield  {journal} {\bibinfo  {journal} {Phys. Rev. Lett.}\
  }\textbf {\bibinfo {volume} {100}},\ \bibinfo {pages} {253001} (\bibinfo
  {year} {2008})}\BibitemShut {NoStop}%
\bibitem [{\citenamefont {Johnson}\ \emph {et~al.}(2008)\citenamefont
  {Johnson}, \citenamefont {Urban}, \citenamefont {Henage}, \citenamefont
  {Isenhower}, \citenamefont {Yavuz}, \citenamefont {Walker},\ and\
  \citenamefont {Saffman}}]{Johnson2008}%
  \BibitemOpen
  \bibfield  {author} {\bibinfo {author} {\bibfnamefont {T.~A.}\ \bibnamefont
  {Johnson}}, \bibinfo {author} {\bibfnamefont {E.}~\bibnamefont {Urban}},
  \bibinfo {author} {\bibfnamefont {T.}~\bibnamefont {Henage}}, \bibinfo
  {author} {\bibfnamefont {L.}~\bibnamefont {Isenhower}}, \bibinfo {author}
  {\bibfnamefont {D.~D.}\ \bibnamefont {Yavuz}}, \bibinfo {author}
  {\bibfnamefont {T.~G.}\ \bibnamefont {Walker}}, \ and\ \bibinfo {author}
  {\bibfnamefont {M.}~\bibnamefont {Saffman}},\ }\href@noop {} {\bibfield
  {journal} {\bibinfo  {journal} {Phys. Rev. Lett.}\ }\textbf {\bibinfo
  {volume} {100}},\ \bibinfo {pages} {113003} (\bibinfo {year}
  {2008})}\BibitemShut {NoStop}%
\bibitem [{\citenamefont {Urban}\ \emph {et~al.}(2009)\citenamefont {Urban},
  \citenamefont {Johnson}, \citenamefont {Henage}, \citenamefont {Isenhower},
  \citenamefont {Yavuz}, \citenamefont {Walker},\ and\ \citenamefont
  {Saffman}}]{Urban2009}%
  \BibitemOpen
  \bibfield  {author} {\bibinfo {author} {\bibfnamefont {E.}~\bibnamefont
  {Urban}}, \bibinfo {author} {\bibfnamefont {T.~A.}\ \bibnamefont {Johnson}},
  \bibinfo {author} {\bibfnamefont {T.}~\bibnamefont {Henage}}, \bibinfo
  {author} {\bibfnamefont {L.}~\bibnamefont {Isenhower}}, \bibinfo {author}
  {\bibfnamefont {D.~D.}\ \bibnamefont {Yavuz}}, \bibinfo {author}
  {\bibfnamefont {T.~G.}\ \bibnamefont {Walker}}, \ and\ \bibinfo {author}
  {\bibfnamefont {M.}~\bibnamefont {Saffman}},\ }\href@noop {} {\bibfield
  {journal} {\bibinfo  {journal} {Nat. Phys.}\ }\textbf {\bibinfo {volume}
  {5}},\ \bibinfo {pages} {110} (\bibinfo {year} {2009})}\BibitemShut {NoStop}%
\bibitem [{\citenamefont {Ga\"etan}\ \emph {et~al.}(2009)\citenamefont
  {Ga\"etan}, \citenamefont {Miroshnychenko}, \citenamefont {Wilk},
  \citenamefont {Chotia}, \citenamefont {Viteau}, \citenamefont {Comparat},
  \citenamefont {Pillet}, \citenamefont {Browaeys},\ and\ \citenamefont
  {Grangier}}]{Gaetan2009}%
  \BibitemOpen
  \bibfield  {author} {\bibinfo {author} {\bibfnamefont {A.}~\bibnamefont
  {Ga\"etan}}, \bibinfo {author} {\bibfnamefont {Y.}~\bibnamefont
  {Miroshnychenko}}, \bibinfo {author} {\bibfnamefont {T.}~\bibnamefont
  {Wilk}}, \bibinfo {author} {\bibfnamefont {A.}~\bibnamefont {Chotia}},
  \bibinfo {author} {\bibfnamefont {M.}~\bibnamefont {Viteau}}, \bibinfo
  {author} {\bibfnamefont {D.}~\bibnamefont {Comparat}}, \bibinfo {author}
  {\bibfnamefont {P.}~\bibnamefont {Pillet}}, \bibinfo {author} {\bibfnamefont
  {A.}~\bibnamefont {Browaeys}}, \ and\ \bibinfo {author} {\bibfnamefont
  {P.}~\bibnamefont {Grangier}},\ }\href@noop {} {\bibfield  {journal}
  {\bibinfo  {journal} {Nat. Phys.}\ }\textbf {\bibinfo {volume} {5}},\
  \bibinfo {pages} {115} (\bibinfo {year} {2009})}\BibitemShut {NoStop}%
\bibitem [{\citenamefont {Saffman}\ \emph {et~al.}(2010)\citenamefont
  {Saffman}, \citenamefont {Walker},\ and\ \citenamefont
  {M\o{}lmer}}]{RevModPhys.82.2313}%
  \BibitemOpen
  \bibfield  {author} {\bibinfo {author} {\bibfnamefont {M.}~\bibnamefont
  {Saffman}}, \bibinfo {author} {\bibfnamefont {T.~G.}\ \bibnamefont {Walker}},
  \ and\ \bibinfo {author} {\bibfnamefont {K.}~\bibnamefont {M\o{}lmer}},\
  }\href@noop {} {\bibfield  {journal} {\bibinfo  {journal} {Rev. Mod. Phys.}\
  }\textbf {\bibinfo {volume} {82}},\ \bibinfo {pages} {2313} (\bibinfo {year}
  {2010})}\BibitemShut {NoStop}%
\bibitem [{\citenamefont {Mayle}\ \emph {et~al.}(2011)\citenamefont {Mayle},
  \citenamefont {Zeller}, \citenamefont {Tezak},\ and\ \citenamefont
  {Schmelcher}}]{Mayle2011}%
  \BibitemOpen
  \bibfield  {author} {\bibinfo {author} {\bibfnamefont {M.}~\bibnamefont
  {Mayle}}, \bibinfo {author} {\bibfnamefont {W.}~\bibnamefont {Zeller}},
  \bibinfo {author} {\bibfnamefont {N.}~\bibnamefont {Tezak}}, \ and\ \bibinfo
  {author} {\bibfnamefont {P.}~\bibnamefont {Schmelcher}},\ }\href@noop {}
  {\bibfield  {journal} {\bibinfo  {journal} {Phys. Rev. A}\ }\textbf {\bibinfo
  {volume} {84}},\ \bibinfo {pages} {010701(R)} (\bibinfo {year}
  {2011})}\BibitemShut {NoStop}%
\bibitem [{\citenamefont {Greene}\ \emph {et~al.}(2000)\citenamefont {Greene},
  \citenamefont {Dickinson},\ and\ \citenamefont {Sadeghpour}}]{greene}%
  \BibitemOpen
  \bibfield  {author} {\bibinfo {author} {\bibfnamefont {C.~H.}\ \bibnamefont
  {Greene}}, \bibinfo {author} {\bibfnamefont {A.~S.}\ \bibnamefont
  {Dickinson}}, \ and\ \bibinfo {author} {\bibfnamefont {H.~R.}\ \bibnamefont
  {Sadeghpour}},\ }\href@noop {} {\bibfield  {journal} {\bibinfo  {journal}
  {Phys. Rev. Lett.}\ }\textbf {\bibinfo {volume} {85}},\ \bibinfo {pages}
  {2458} (\bibinfo {year} {2000})}\BibitemShut {NoStop}%
\bibitem [{\citenamefont {Bendkowsky}\ \emph {et~al.}(2009)\citenamefont
  {Bendkowsky}, \citenamefont {Butscher}, \citenamefont {Nipper}, \citenamefont
  {Shaffer}, \citenamefont {Low},\ and\ \citenamefont {Pfau}}]{Bendkowsky2009}%
  \BibitemOpen
  \bibfield  {author} {\bibinfo {author} {\bibfnamefont {V.}~\bibnamefont
  {Bendkowsky}}, \bibinfo {author} {\bibfnamefont {B.}~\bibnamefont
  {Butscher}}, \bibinfo {author} {\bibfnamefont {J.}~\bibnamefont {Nipper}},
  \bibinfo {author} {\bibfnamefont {J.~P.}\ \bibnamefont {Shaffer}}, \bibinfo
  {author} {\bibfnamefont {R.}~\bibnamefont {Low}}, \ and\ \bibinfo {author}
  {\bibfnamefont {T.}~\bibnamefont {Pfau}},\ }\href@noop {} {\bibfield
  {journal} {\bibinfo  {journal} {Nature}\ }\textbf {\bibinfo {volume} {458}},\
  \bibinfo {pages} {1005} (\bibinfo {year} {2009})}\BibitemShut {NoStop}%
\bibitem [{\citenamefont {Bendkowsky}\ \emph {et~al.}(2010)\citenamefont
  {Bendkowsky}, \citenamefont {Butscher}, \citenamefont {Nipper}, \citenamefont
  {Balewski}, \citenamefont {Shaffer}, \citenamefont {L\"ow}, \citenamefont
  {Pfau}, \citenamefont {Li}, \citenamefont {Stanojevic}, \citenamefont
  {Pohl},\ and\ \citenamefont {Rost}}]{PhysRevLett.105.163201}%
  \BibitemOpen
  \bibfield  {author} {\bibinfo {author} {\bibfnamefont {V.}~\bibnamefont
  {Bendkowsky}}, \bibinfo {author} {\bibfnamefont {B.}~\bibnamefont
  {Butscher}}, \bibinfo {author} {\bibfnamefont {J.}~\bibnamefont {Nipper}},
  \bibinfo {author} {\bibfnamefont {J.~B.}\ \bibnamefont {Balewski}}, \bibinfo
  {author} {\bibfnamefont {J.~P.}\ \bibnamefont {Shaffer}}, \bibinfo {author}
  {\bibfnamefont {R.}~\bibnamefont {L\"ow}}, \bibinfo {author} {\bibfnamefont
  {T.}~\bibnamefont {Pfau}}, \bibinfo {author} {\bibfnamefont {W.}~\bibnamefont
  {Li}}, \bibinfo {author} {\bibfnamefont {J.}~\bibnamefont {Stanojevic}},
  \bibinfo {author} {\bibfnamefont {T.}~\bibnamefont {Pohl}}, \ and\ \bibinfo
  {author} {\bibfnamefont {J.~M.}\ \bibnamefont {Rost}},\ }\href@noop {}
  {\bibfield  {journal} {\bibinfo  {journal} {Phys. Rev. Lett.}\ }\textbf
  {\bibinfo {volume} {105}},\ \bibinfo {pages} {163201} (\bibinfo {year}
  {2010})}\BibitemShut {NoStop}%
\bibitem [{\citenamefont {Butscher}\ \emph {et~al.}(2010)\citenamefont
  {Butscher}, \citenamefont {Nipper}, \citenamefont {Balewski}, \citenamefont
  {Kukota}, \citenamefont {Bendkowsky}, \citenamefont {Low},\ and\
  \citenamefont {Pfau}}]{Butscher2010}%
  \BibitemOpen
  \bibfield  {author} {\bibinfo {author} {\bibfnamefont {B.}~\bibnamefont
  {Butscher}}, \bibinfo {author} {\bibfnamefont {J.}~\bibnamefont {Nipper}},
  \bibinfo {author} {\bibfnamefont {J.~B.}\ \bibnamefont {Balewski}}, \bibinfo
  {author} {\bibfnamefont {L.}~\bibnamefont {Kukota}}, \bibinfo {author}
  {\bibfnamefont {V.}~\bibnamefont {Bendkowsky}}, \bibinfo {author}
  {\bibfnamefont {R.}~\bibnamefont {Low}}, \ and\ \bibinfo {author}
  {\bibfnamefont {T.}~\bibnamefont {Pfau}},\ }\href@noop {} {\bibfield
  {journal} {\bibinfo  {journal} {Nat. Phys.}\ }\textbf {\bibinfo {volume}
  {6}},\ \bibinfo {pages} {970} (\bibinfo {year} {2010})}\BibitemShut {NoStop}%
\bibitem [{\citenamefont {Rittenhouse}\ and\ \citenamefont
  {Sadeghpour}(2010)}]{PhysRevLett.104.243002}%
  \BibitemOpen
  \bibfield  {author} {\bibinfo {author} {\bibfnamefont {S.~T.}\ \bibnamefont
  {Rittenhouse}}\ and\ \bibinfo {author} {\bibfnamefont {H.~R.}\ \bibnamefont
  {Sadeghpour}},\ }\href@noop {} {\bibfield  {journal} {\bibinfo  {journal}
  {Phys. Rev. Lett.}\ }\textbf {\bibinfo {volume} {104}},\ \bibinfo {pages}
  {243002} (\bibinfo {year} {2010})}\BibitemShut {NoStop}%
\bibitem [{\citenamefont {Rittenhouse}\ \emph {et~al.}(2011)\citenamefont
  {Rittenhouse}, \citenamefont {Mayle}, \citenamefont {Schmelcher},\ and\
  \citenamefont {Sadeghpour}}]{Rittenhouse2011a}%
  \BibitemOpen
  \bibfield  {author} {\bibinfo {author} {\bibfnamefont {S.~T.}\ \bibnamefont
  {Rittenhouse}}, \bibinfo {author} {\bibfnamefont {M.}~\bibnamefont {Mayle}},
  \bibinfo {author} {\bibfnamefont {P.}~\bibnamefont {Schmelcher}}, \ and\
  \bibinfo {author} {\bibfnamefont {H.~R.}\ \bibnamefont {Sadeghpour}},\
  }\href@noop {} {\bibfield  {journal} {\bibinfo  {journal} {J. Phys. B}\
  }\textbf {\bibinfo {volume} {44}},\ \bibinfo {pages} {184005} (\bibinfo
  {year} {2011})}\BibitemShut {NoStop}%
\bibitem [{\citenamefont {Hyafil}\ \emph {et~al.}(2004)\citenamefont {Hyafil},
  \citenamefont {Mozley}, \citenamefont {Perrin}, \citenamefont {Tailleur},
  \citenamefont {Nogues}, \citenamefont {Brune}, \citenamefont {Raimond},\ and\
  \citenamefont {Haroche}}]{Hyafil2004}%
  \BibitemOpen
  \bibfield  {author} {\bibinfo {author} {\bibfnamefont {P.}~\bibnamefont
  {Hyafil}}, \bibinfo {author} {\bibfnamefont {J.}~\bibnamefont {Mozley}},
  \bibinfo {author} {\bibfnamefont {A.}~\bibnamefont {Perrin}}, \bibinfo
  {author} {\bibfnamefont {J.}~\bibnamefont {Tailleur}}, \bibinfo {author}
  {\bibfnamefont {G.}~\bibnamefont {Nogues}}, \bibinfo {author} {\bibfnamefont
  {M.}~\bibnamefont {Brune}}, \bibinfo {author} {\bibfnamefont {J.~M.}\
  \bibnamefont {Raimond}}, \ and\ \bibinfo {author} {\bibfnamefont
  {S.}~\bibnamefont {Haroche}},\ }\href@noop {} {\bibfield  {journal} {\bibinfo
   {journal} {Phys. Rev. Lett.}\ }\textbf {\bibinfo {volume} {93}},\ \bibinfo
  {pages} {103001} (\bibinfo {year} {2004})}\BibitemShut {NoStop}%
\bibitem [{\citenamefont {Dutta}\ \emph {et~al.}(2000)\citenamefont {Dutta},
  \citenamefont {Guest}, \citenamefont {Feldbaum}, \citenamefont
  {Walz-Flannigan},\ and\ \citenamefont {Raithel}}]{Dutta2000}%
  \BibitemOpen
  \bibfield  {author} {\bibinfo {author} {\bibfnamefont {S.~K.}\ \bibnamefont
  {Dutta}}, \bibinfo {author} {\bibfnamefont {J.~R.}\ \bibnamefont {Guest}},
  \bibinfo {author} {\bibfnamefont {D.}~\bibnamefont {Feldbaum}}, \bibinfo
  {author} {\bibfnamefont {A.}~\bibnamefont {Walz-Flannigan}}, \ and\ \bibinfo
  {author} {\bibfnamefont {G.}~\bibnamefont {Raithel}},\ }\href@noop {}
  {\bibfield  {journal} {\bibinfo  {journal} {Phys. Rev. Lett.}\ }\textbf
  {\bibinfo {volume} {85}},\ \bibinfo {pages} {5551} (\bibinfo {year}
  {2000})}\BibitemShut {NoStop}%
\bibitem [{\citenamefont {Younge}\ \emph
  {et~al.}(2010{\natexlab{a}})\citenamefont {Younge}, \citenamefont
  {Anderson},\ and\ \citenamefont {Raithel}}]{1367-2630-12-2-023031}%
  \BibitemOpen
  \bibfield  {author} {\bibinfo {author} {\bibfnamefont {K.~C.}\ \bibnamefont
  {Younge}}, \bibinfo {author} {\bibfnamefont {S.~E.}\ \bibnamefont
  {Anderson}}, \ and\ \bibinfo {author} {\bibfnamefont {G.}~\bibnamefont
  {Raithel}},\ }\href@noop {} {\bibfield  {journal} {\bibinfo  {journal} {New
  J. Phys.}\ }\textbf {\bibinfo {volume} {12}},\ \bibinfo {pages} {023031}
  (\bibinfo {year} {2010}{\natexlab{a}})}\BibitemShut {NoStop}%
\bibitem [{\citenamefont {Younge}\ \emph
  {et~al.}(2010{\natexlab{b}})\citenamefont {Younge}, \citenamefont {Knuffman},
  \citenamefont {Anderson},\ and\ \citenamefont
  {Raithel}}]{PhysRevLett.104.173001}%
  \BibitemOpen
  \bibfield  {author} {\bibinfo {author} {\bibfnamefont {K.~C.}\ \bibnamefont
  {Younge}}, \bibinfo {author} {\bibfnamefont {B.}~\bibnamefont {Knuffman}},
  \bibinfo {author} {\bibfnamefont {S.~E.}\ \bibnamefont {Anderson}}, \ and\
  \bibinfo {author} {\bibfnamefont {G.}~\bibnamefont {Raithel}},\ }\href@noop
  {} {\bibfield  {journal} {\bibinfo  {journal} {Phys. Rev. Lett.}\ }\textbf
  {\bibinfo {volume} {104}},\ \bibinfo {pages} {173001} (\bibinfo {year}
  {2010}{\natexlab{b}})}\BibitemShut {NoStop}%
\bibitem [{\citenamefont {Choi}\ \emph {et~al.}(2005)\citenamefont {Choi},
  \citenamefont {Guest}, \citenamefont {Povilus}, \citenamefont {Hansis},\ and\
  \citenamefont {Raithel}}]{choi}%
  \BibitemOpen
  \bibfield  {author} {\bibinfo {author} {\bibfnamefont {J.-H.}\ \bibnamefont
  {Choi}}, \bibinfo {author} {\bibfnamefont {J.~R.}\ \bibnamefont {Guest}},
  \bibinfo {author} {\bibfnamefont {A.~P.}\ \bibnamefont {Povilus}}, \bibinfo
  {author} {\bibfnamefont {E.}~\bibnamefont {Hansis}}, \ and\ \bibinfo {author}
  {\bibfnamefont {G.}~\bibnamefont {Raithel}},\ }\href@noop {} {\bibfield
  {journal} {\bibinfo  {journal} {Phys. Rev. Lett.}\ }\textbf {\bibinfo
  {volume} {95}},\ \bibinfo {pages} {243001} (\bibinfo {year}
  {2005})}\BibitemShut {NoStop}%
\bibitem [{\citenamefont {Lesanovsky}\ and\ \citenamefont
  {Schmelcher}(2005)}]{LesanovskyPRL2005}%
  \BibitemOpen
  \bibfield  {author} {\bibinfo {author} {\bibfnamefont {I.}~\bibnamefont
  {Lesanovsky}}\ and\ \bibinfo {author} {\bibfnamefont {P.}~\bibnamefont
  {Schmelcher}},\ }\href@noop {} {\bibfield  {journal} {\bibinfo  {journal}
  {Phys. Rev. Lett.}\ }\textbf {\bibinfo {volume} {95}},\ \bibinfo {pages}
  {053001} (\bibinfo {year} {2005})}\BibitemShut {NoStop}%
\bibitem [{\citenamefont {Hezel}\ \emph {et~al.}(2006)\citenamefont {Hezel},
  \citenamefont {Lesanovsky},\ and\ \citenamefont {Schmelcher}}]{Hezel2006}%
  \BibitemOpen
  \bibfield  {author} {\bibinfo {author} {\bibfnamefont {B.}~\bibnamefont
  {Hezel}}, \bibinfo {author} {\bibfnamefont {I.}~\bibnamefont {Lesanovsky}}, \
  and\ \bibinfo {author} {\bibfnamefont {P.}~\bibnamefont {Schmelcher}},\
  }\href@noop {} {\bibfield  {journal} {\bibinfo  {journal} {Phys. Rev. Lett.}\
  }\textbf {\bibinfo {volume} {97}},\ \bibinfo {pages} {223001} (\bibinfo
  {year} {2006})}\BibitemShut {NoStop}%
\bibitem [{\citenamefont {Mayle}\ \emph {et~al.}(2009)\citenamefont {Mayle},
  \citenamefont {Lesanovsky},\ and\ \citenamefont {Schmelcher}}]{mayle:053410}%
  \BibitemOpen
  \bibfield  {author} {\bibinfo {author} {\bibfnamefont {M.}~\bibnamefont
  {Mayle}}, \bibinfo {author} {\bibfnamefont {I.}~\bibnamefont {Lesanovsky}}, \
  and\ \bibinfo {author} {\bibfnamefont {P.}~\bibnamefont {Schmelcher}},\
  }\href@noop {} {\bibfield  {journal} {\bibinfo  {journal} {Phys. Rev. A}\
  }\textbf {\bibinfo {volume} {80}},\ \bibinfo {pages} {053410} (\bibinfo
  {year} {2009})}\BibitemShut {NoStop}%
\bibitem [{\citenamefont {Mayle}\ \emph {et~al.}(2007)\citenamefont {Mayle},
  \citenamefont {Hezel}, \citenamefont {Lesanovsky},\ and\ \citenamefont
  {Schmelcher}}]{mayle:113004}%
  \BibitemOpen
  \bibfield  {author} {\bibinfo {author} {\bibfnamefont {M.}~\bibnamefont
  {Mayle}}, \bibinfo {author} {\bibfnamefont {B.}~\bibnamefont {Hezel}},
  \bibinfo {author} {\bibfnamefont {I.}~\bibnamefont {Lesanovsky}}, \ and\
  \bibinfo {author} {\bibfnamefont {P.}~\bibnamefont {Schmelcher}},\
  }\href@noop {} {\bibfield  {journal} {\bibinfo  {journal} {Phys. Rev. Lett.}\
  }\textbf {\bibinfo {volume} {99}},\ \bibinfo {pages} {113004} (\bibinfo
  {year} {2007})}\BibitemShut {NoStop}%
\bibitem [{\citenamefont {Hezel}\ \emph {et~al.}(2007)\citenamefont {Hezel},
  \citenamefont {Lesanovsky},\ and\ \citenamefont {Schmelcher}}]{HezelPRA}%
  \BibitemOpen
  \bibfield  {author} {\bibinfo {author} {\bibfnamefont {B.}~\bibnamefont
  {Hezel}}, \bibinfo {author} {\bibfnamefont {I.}~\bibnamefont {Lesanovsky}}, \
  and\ \bibinfo {author} {\bibfnamefont {P.}~\bibnamefont {Schmelcher}},\
  }\href@noop {} {\bibfield  {journal} {\bibinfo  {journal} {Phys. Rev. A}\
  }\textbf {\bibinfo {volume} {76}},\ \bibinfo {pages} {053417} (\bibinfo
  {year} {2007})}\BibitemShut {NoStop}%
\bibitem [{\citenamefont {Pritchard}(1983)}]{PhysRevLett.51.1336}%
  \BibitemOpen
  \bibfield  {author} {\bibinfo {author} {\bibfnamefont {D.~E.}\ \bibnamefont
  {Pritchard}},\ }\href@noop {} {\bibfield  {journal} {\bibinfo  {journal}
  {Phys. Rev. Lett.}\ }\textbf {\bibinfo {volume} {51}},\ \bibinfo {pages}
  {1336} (\bibinfo {year} {1983})}\BibitemShut {NoStop}%
\bibitem [{\citenamefont {Moore}\ \emph {et~al.}(2006)\citenamefont {Moore},
  \citenamefont {Purdy}, \citenamefont {Murch}, \citenamefont {Brown},
  \citenamefont {Dani}, \citenamefont {Gupta},\ and\ \citenamefont
  {Stamper-Kurn}}]{Moore}%
  \BibitemOpen
  \bibfield  {author} {\bibinfo {author} {\bibfnamefont {K.~L.}\ \bibnamefont
  {Moore}}, \bibinfo {author} {\bibfnamefont {T.~P.}\ \bibnamefont {Purdy}},
  \bibinfo {author} {\bibfnamefont {K.~W.}\ \bibnamefont {Murch}}, \bibinfo
  {author} {\bibfnamefont {K.~R.}\ \bibnamefont {Brown}}, \bibinfo {author}
  {\bibfnamefont {K.}~\bibnamefont {Dani}}, \bibinfo {author} {\bibfnamefont
  {S.}~\bibnamefont {Gupta}}, \ and\ \bibinfo {author} {\bibfnamefont {D.~M.}\
  \bibnamefont {Stamper-Kurn}},\ }\href@noop {} {\bibfield  {journal} {\bibinfo
   {journal} {Appl. Phys. B}\ }\textbf {\bibinfo {volume} {82}},\ \bibinfo
  {pages} {533} (\bibinfo {year} {2006})}\BibitemShut {NoStop}%
\bibitem [{\citenamefont {Dalgarno}\ and\ \citenamefont
  {Davison}(1966)}]{Dalgarno1966}%
  \BibitemOpen
  \bibfield  {author} {\bibinfo {author} {\bibfnamefont {A.}~\bibnamefont
  {Dalgarno}}\ and\ \bibinfo {author} {\bibfnamefont {W.~D.}\ \bibnamefont
  {Davison}},\ }\href@noop {} {\bibfield  {journal} {\bibinfo  {journal}
  {Advances in Atomic and Molecular Physics}\ }\textbf {\bibinfo {volume}
  {2}},\ \bibinfo {pages} {1} (\bibinfo {year} {1966})}\BibitemShut {NoStop}%
\bibitem [{\citenamefont {Jentschura}\ \emph {et~al.}(2005)\citenamefont
  {Jentschura}, \citenamefont {Bigot}, \citenamefont {Evers}, \citenamefont
  {Mohr},\ and\ \citenamefont {Keitel}}]{Jentschura2005}%
  \BibitemOpen
  \bibfield  {author} {\bibinfo {author} {\bibfnamefont {U.~D.}\ \bibnamefont
  {Jentschura}}, \bibinfo {author} {\bibfnamefont {E.-O.~L.}\ \bibnamefont
  {Bigot}}, \bibinfo {author} {\bibfnamefont {J.}~\bibnamefont {Evers}},
  \bibinfo {author} {\bibfnamefont {P.~J.}\ \bibnamefont {Mohr}}, \ and\
  \bibinfo {author} {\bibfnamefont {C.~H.}\ \bibnamefont {Keitel}},\
  }\href@noop {} {\bibfield  {journal} {\bibinfo  {journal} {J. Phys. B}\
  }\textbf {\bibinfo {volume} {38}},\ \bibinfo {pages} {S97} (\bibinfo {year}
  {2005})}\BibitemShut {NoStop}%
\bibitem [{\citenamefont {Folman}\ \emph {et~al.}(2002)\citenamefont {Folman},
  \citenamefont {Kr\"uger}, \citenamefont {Schmiedmayer}, \citenamefont
  {Denschlag},\ and\ \citenamefont {Henkel}}]{folman}%
  \BibitemOpen
  \bibfield  {author} {\bibinfo {author} {\bibfnamefont {R.}~\bibnamefont
  {Folman}}, \bibinfo {author} {\bibfnamefont {P.}~\bibnamefont {Kr\"uger}},
  \bibinfo {author} {\bibfnamefont {J.}~\bibnamefont {Schmiedmayer}}, \bibinfo
  {author} {\bibfnamefont {J.}~\bibnamefont {Denschlag}}, \ and\ \bibinfo
  {author} {\bibfnamefont {C.}~\bibnamefont {Henkel}},\ }\href@noop {}
  {\bibfield  {journal} {\bibinfo  {journal} {Adv.~At.~Mol.~Opt.~Phys.}\
  }\textbf {\bibinfo {volume} {48}},\ \bibinfo {pages} {263} (\bibinfo {year}
  {2002})}\BibitemShut {NoStop}%
\bibitem [{\citenamefont {Hulet}\ and\ \citenamefont
  {Kleppner}(1983)}]{Hulet1983}%
  \BibitemOpen
  \bibfield  {author} {\bibinfo {author} {\bibfnamefont {R.~G.}\ \bibnamefont
  {Hulet}}\ and\ \bibinfo {author} {\bibfnamefont {D.}~\bibnamefont
  {Kleppner}},\ }\href@noop {} {\bibfield  {journal} {\bibinfo  {journal}
  {Phys. Rev. Lett.}\ }\textbf {\bibinfo {volume} {51}},\ \bibinfo {pages}
  {1430} (\bibinfo {year} {1983})}\BibitemShut {NoStop}%
\bibitem [{\citenamefont {Delande}\ and\ \citenamefont
  {Gay}(1988)}]{Delande1988}%
  \BibitemOpen
  \bibfield  {author} {\bibinfo {author} {\bibfnamefont {D.}~\bibnamefont
  {Delande}}\ and\ \bibinfo {author} {\bibfnamefont {J.~C.}\ \bibnamefont
  {Gay}},\ }\href@noop {} {\bibfield  {journal} {\bibinfo  {journal} {Europhys.
  Lett.)}\ }\textbf {\bibinfo {volume} {5}},\ \bibinfo {pages} {303} (\bibinfo
  {year} {1988})}\BibitemShut {NoStop}%
\bibitem [{\citenamefont {Molander}\ \emph {et~al.}(1986)\citenamefont
  {Molander}, \citenamefont {Stroud},\ and\ \citenamefont
  {Yeazell}}]{Molander1986}%
  \BibitemOpen
  \bibfield  {author} {\bibinfo {author} {\bibfnamefont {W.~A.}\ \bibnamefont
  {Molander}}, \bibinfo {author} {\bibfnamefont {C.~R.}\ \bibnamefont
  {Stroud}}, \ and\ \bibinfo {author} {\bibfnamefont {J.~A.}\ \bibnamefont
  {Yeazell}},\ }\href@noop {} {\bibfield  {journal} {\bibinfo  {journal} {J.
  Phys. B}\ }\textbf {\bibinfo {volume} {19}},\ \bibinfo {pages} {L461}
  (\bibinfo {year} {1986})}\BibitemShut {NoStop}%
\bibitem [{\citenamefont {{Ashkin}}\ \emph {et~al.}(1986)\citenamefont
  {{Ashkin}}, \citenamefont {{Dziedzic}}, \citenamefont {{Bjorkholm}},\ and\
  \citenamefont {{Chu}}}]{Ashkin1986}%
  \BibitemOpen
  \bibfield  {author} {\bibinfo {author} {\bibfnamefont {A.}~\bibnamefont
  {{Ashkin}}}, \bibinfo {author} {\bibfnamefont {J.~M.}\ \bibnamefont
  {{Dziedzic}}}, \bibinfo {author} {\bibfnamefont {J.~E.}\ \bibnamefont
  {{Bjorkholm}}}, \ and\ \bibinfo {author} {\bibfnamefont {S.}~\bibnamefont
  {{Chu}}},\ }\href@noop {} {\bibfield  {journal} {\bibinfo  {journal} {Opt.
  Lett.}\ }\textbf {\bibinfo {volume} {11}},\ \bibinfo {pages} {288} (\bibinfo
  {year} {1986})}\BibitemShut {NoStop}%
\bibitem [{\citenamefont {Schlosser}\ \emph {et~al.}(2001)\citenamefont
  {Schlosser}, \citenamefont {Reymond}, \citenamefont {Protsenko},\ and\
  \citenamefont {Grangier}}]{Schlosser2001}%
  \BibitemOpen
  \bibfield  {author} {\bibinfo {author} {\bibfnamefont {N.}~\bibnamefont
  {Schlosser}}, \bibinfo {author} {\bibfnamefont {G.}~\bibnamefont {Reymond}},
  \bibinfo {author} {\bibfnamefont {I.}~\bibnamefont {Protsenko}}, \ and\
  \bibinfo {author} {\bibfnamefont {P.}~\bibnamefont {Grangier}},\ }\href@noop
  {} {\bibfield  {journal} {\bibinfo  {journal} {Nature}\ }\textbf {\bibinfo
  {volume} {411}},\ \bibinfo {pages} {1024} (\bibinfo {year}
  {2001})}\BibitemShut {NoStop}%
\bibitem [{\citenamefont {Vuletic}(2006)}]{Vuletic2006}%
  \BibitemOpen
  \bibfield  {author} {\bibinfo {author} {\bibfnamefont {V.}~\bibnamefont
  {Vuletic}},\ }\href@noop {} {\bibfield  {journal} {\bibinfo  {journal} {Nat.
  Phys.}\ }\textbf {\bibinfo {volume} {2}},\ \bibinfo {pages} {801} (\bibinfo
  {year} {2006})}\BibitemShut {NoStop}%
\bibitem [{\citenamefont {Nussenzveig}\ \emph {et~al.}(1993)\citenamefont
  {Nussenzveig}, \citenamefont {Bernardot}, \citenamefont {Brune},
  \citenamefont {Hare}, \citenamefont {Raimond}, \citenamefont {Haroche},\ and\
  \citenamefont {Gawlik}}]{Nussenzveig1993}%
  \BibitemOpen
  \bibfield  {author} {\bibinfo {author} {\bibfnamefont {P.}~\bibnamefont
  {Nussenzveig}}, \bibinfo {author} {\bibfnamefont {F.}~\bibnamefont
  {Bernardot}}, \bibinfo {author} {\bibfnamefont {M.}~\bibnamefont {Brune}},
  \bibinfo {author} {\bibfnamefont {J.}~\bibnamefont {Hare}}, \bibinfo {author}
  {\bibfnamefont {J.~M.}\ \bibnamefont {Raimond}}, \bibinfo {author}
  {\bibfnamefont {S.}~\bibnamefont {Haroche}}, \ and\ \bibinfo {author}
  {\bibfnamefont {W.}~\bibnamefont {Gawlik}},\ }\href@noop {} {\bibfield
  {journal} {\bibinfo  {journal} {Phys. Rev. A}\ }\textbf {\bibinfo {volume}
  {48}},\ \bibinfo {pages} {3991} (\bibinfo {year} {1993})}\BibitemShut
  {NoStop}%
\bibitem [{\citenamefont {Pohl}\ \emph {et~al.}(2010)\citenamefont {Pohl},
  \citenamefont {Demler},\ and\ \citenamefont {Lukin}}]{Pohl2010}%
  \BibitemOpen
  \bibfield  {author} {\bibinfo {author} {\bibfnamefont {T.}~\bibnamefont
  {Pohl}}, \bibinfo {author} {\bibfnamefont {E.}~\bibnamefont {Demler}}, \ and\
  \bibinfo {author} {\bibfnamefont {M.~D.}\ \bibnamefont {Lukin}},\ }\href@noop
  {} {\bibfield  {journal} {\bibinfo  {journal} {Phys. Rev. Lett.}\ }\textbf
  {\bibinfo {volume} {104}},\ \bibinfo {pages} {043002} (\bibinfo {year}
  {2010})}\BibitemShut {NoStop}%
\bibitem [{\citenamefont {Cheng}\ \emph {et~al.}(1994)\citenamefont {Cheng},
  \citenamefont {Lee},\ and\ \citenamefont {Gallagher}}]{Cheng1994}%
  \BibitemOpen
  \bibfield  {author} {\bibinfo {author} {\bibfnamefont {C.~H.}\ \bibnamefont
  {Cheng}}, \bibinfo {author} {\bibfnamefont {C.~Y.}\ \bibnamefont {Lee}}, \
  and\ \bibinfo {author} {\bibfnamefont {T.~F.}\ \bibnamefont {Gallagher}},\
  }\href@noop {} {\bibfield  {journal} {\bibinfo  {journal} {Phys. Rev. Lett.}\
  }\textbf {\bibinfo {volume} {73}},\ \bibinfo {pages} {3078} (\bibinfo {year}
  {1994})}\BibitemShut {NoStop}%
\bibitem [{\citenamefont {Tezak}\ \emph {et~al.}()\citenamefont {Tezak},
  \citenamefont {Mayle},\ and\ \citenamefont {Schmelcher}}]{Tezak2010}%
  \BibitemOpen
  \bibfield  {author} {\bibinfo {author} {\bibfnamefont {N.}~\bibnamefont
  {Tezak}}, \bibinfo {author} {\bibfnamefont {M.}~\bibnamefont {Mayle}}, \ and\
  \bibinfo {author} {\bibfnamefont {P.}~\bibnamefont {Schmelcher}},\
  }\href@noop {} {\bibinfo  {journal} {arXiv:1012.3810v1 [physics.atom-ph]}\
  }\BibitemShut {NoStop}%
\bibitem [{\citenamefont {Messiah}(1961)}]{Messiah2}%
  \BibitemOpen
\bibfield  {journal} {  }\bibfield  {author} {\bibinfo {author} {\bibfnamefont
  {A.}~\bibnamefont {Messiah}},\ }\href@noop {} {\emph {\bibinfo {title}
  {Quantum Mechanics}}},\ Vol.~\bibinfo {volume} {II}\ (\bibinfo  {publisher}
  {North-Holland Publishing Company},\ \bibinfo {year} {1961})\BibitemShut
  {NoStop}%
\bibitem [{\citenamefont {Zimmerman}\ \emph {et~al.}(1979)\citenamefont
  {Zimmerman}, \citenamefont {Littman}, \citenamefont {Kash},\ and\
  \citenamefont {Kleppner}}]{Zimmermann1979}%
  \BibitemOpen
  \bibfield  {author} {\bibinfo {author} {\bibfnamefont {M.~L.}\ \bibnamefont
  {Zimmerman}}, \bibinfo {author} {\bibfnamefont {M.~G.}\ \bibnamefont
  {Littman}}, \bibinfo {author} {\bibfnamefont {M.~M.}\ \bibnamefont {Kash}}, \
  and\ \bibinfo {author} {\bibfnamefont {D.}~\bibnamefont {Kleppner}},\
  }\href@noop {} {\bibfield  {journal} {\bibinfo  {journal} {Phys. Rev. A}\
  }\textbf {\bibinfo {volume} {20}},\ \bibinfo {pages} {2251} (\bibinfo {year}
  {1979})}\BibitemShut {NoStop}%
\end{thebibliography}%

\end{document}